\documentclass[twocolumn]{aastex631}

\newcommand\tess{TESS}
\newcommand\gaia{\textit{Gaia}}
\newcommand\kms{$\textrm{km~s}^{-1}$}
\newcommand\ms{$\textrm{m~s}^{-1}$}
\newcommand\cms{$\textrm{cm~s}^{-1}$}
\newcommand\gcmcubed{$\textrm{g~cm}^{-3}$}

\newcommand\teff{T$_{\rm{eff}}$}
\newcommand\vsini{$v$~sin~$i$}

\newcommand{\unit}[1]{\ensuremath{\, \mathrm{#1}}} 
\newcommand\earthmass{$M_{\oplus}$}
\newcommand\earthradius{$R_{\oplus}$}
\newcommand\solmass{$M_{\odot}$}

\newcommand\plmass{85.3$^{+8.8}_{-8.7}$}
\newcommand\plradius{12.0$^{+0.4}_{-0.5}$}

\shortauthors{Kanodia et al. 2022}
\shorttitle{An inflated Jovian planet around TOI-3757}

\begin{document}

\title{TOI-3757 b: A low density gas giant orbiting a solar-metallicity M dwarf}

\author[0000-0001-8401-4300]{Shubham Kanodia}
\affil{Department of Astronomy \& Astrophysics, 525 Davey Laboratory, The Pennsylvania State University, University Park, PA, 16802, USA}
\affil{Center for Exoplanets and Habitable Worlds, 525 Davey Laboratory, The Pennsylvania State University, University Park, PA, 16802, USA}

\author[0000-0002-2990-7613]{Jessica Libby-Roberts}
\affil{Department of Astronomy \& Astrophysics, 525 Davey Laboratory, The Pennsylvania State University, University Park, PA, 16802, USA}
\affil{Center for Exoplanets and Habitable Worlds, 525 Davey Laboratory, The Pennsylvania State University, University Park, PA, 16802, USA}

\author[0000-0003-4835-0619]{Caleb I. Ca\~nas}
\affil{Department of Astronomy \& Astrophysics, 525 Davey Laboratory, The Pennsylvania State University, University Park, PA, 16802, USA}
\affil{Center for Exoplanets and Habitable Worlds, 525 Davey Laboratory, The Pennsylvania State University, University Park, PA, 16802, USA}
\affil{NASA Earth and Space Science Fellow}

\author[0000-0001-8720-5612]{Joe P. Ninan}
\affil{Department of Astronomy \& Astrophysics, 525 Davey Laboratory, The Pennsylvania State University, University Park, PA, 16802, USA}
\affil{Center for Exoplanets and Habitable Worlds, 525 Davey Laboratory, The Pennsylvania State University, University Park, PA, 16802, USA}

\author[0000-0001-9596-7983]{Suvrath Mahadevan}
\affil{Department of Astronomy \& Astrophysics, 525 Davey Laboratory, The Pennsylvania State University, University Park, PA, 16802, USA}
\affil{Center for Exoplanets and Habitable Worlds, 525 Davey Laboratory, The Pennsylvania State University, University Park, PA, 16802, USA}

\author[0000-0001-7409-5688]{Gudmundur Stefansson}
\affil{Henry Norris Russell Fellow}
\affil{Department of Astrophysical Sciences, Princeton University, 4 Ivy Lane, Princeton, NJ 08540, USA}

\author[0000-0002-9082-6337]{Andrea S.J. Lin}
\affil{Department of Astronomy \& Astrophysics, 525 Davey Laboratory, The Pennsylvania State University, University Park, PA, 16802, USA}
\affil{Center for Exoplanets and Habitable Worlds, 525 Davey Laboratory, The Pennsylvania State University, University Park, PA, 16802, USA}

\author[0000-0002-7227-2334]{Sinclaire Jones}
\affil{Department of Astrophysical Sciences, Princeton University, 4 Ivy Lane, Princeton, NJ 08540, USA}

\author[0000-0002-0048-2586]{Andrew Monson}
\affil{Department of Astronomy \& Astrophysics, 525 Davey Laboratory, The Pennsylvania State University, University Park, PA, 16802, USA}
\affil{Center for Exoplanets and Habitable Worlds, 525 Davey Laboratory, The Pennsylvania State University, University Park, PA, 16802, USA}

\author[0000-0001-9307-8170]{Brock A. Parker}
\affil{Department of Physics \& Astronomy, University of Wyoming, Laramie, WY 82070, USA}

\author[0000-0002-4475-4176]{Henry A. Kobulnicky}
\affil{Department of Physics \& Astronomy, University of Wyoming, Laramie, WY 82070, USA}

\author[0000-0002-5817-202X]{Tera N. Swaby}
\affil{Department of Physics \& Astronomy, University of Wyoming, Laramie, WY 82070, USA}

\author[0000-0002-5300-5353]{Luke Powers}
\affil{Department of Astronomy \& Astrophysics, 525 Davey Laboratory, The Pennsylvania State University, University Park, PA, 16802, USA}
\affil{Center for Exoplanets and Habitable Worlds, 525 Davey Laboratory, The Pennsylvania State University, University Park, PA, 16802, USA}

\author[0000-0001-7708-2364]{Corey Beard}
\affil{Department of Physics and Astronomy, The University of California, Irvine, Irvine, CA 92697, USA}

\author[0000-0003-4384-7220]{Chad F. Bender}
\affil{Steward Observatory, The University of Arizona, 933 N.\ Cherry Avenue, Tucson, AZ 85721, USA}

\author[0000-0002-6096-1749]{Cullen H.\ Blake}
\affil{Department of Physics and Astronomy, University of Pennsylvania, 209 S 33rd St, Philadelphia, PA 19104, USA}

\author[0000-0001-9662-3496]{William D. Cochran}
\affil{McDonald Observatory and Department of Astronomy, The University of Texas at Austin, USA}
\affil{Center for Planetary Systems Habitability, The University of Texas at Austin, USA}

\author[0000-0002-3610-6953]{Jiayin Dong}
\affil{Department of Astronomy \& Astrophysics, 525 Davey Laboratory, The Pennsylvania State University, University Park, PA, 16802, USA}
\affil{Center for Exoplanets and Habitable Worlds, 525 Davey Laboratory, The Pennsylvania State University, University Park, PA, 16802, USA}

\author[0000-0002-2144-0764]{Scott A. Diddams}
\affil{Time and Frequency Division, National Institute of Standards and Technology, 325 Broadway, Boulder, CO 80305, USA}
\affil{Department of Physics, University of Colorado, 2000 Colorado Avenue, Boulder, CO 80309, USA}
\affil{Electrical, Computer \& Energy Engineering, University of Colorado, 425 UCB, Boulder, CO 80309, USA}

\author[0000-0002-0560-1433]{Connor Fredrick}
\affil{Time and Frequency Division, National Institute of Standards and Technology, 325 Broadway, Boulder, CO 80305, USA}
\affil{Department of Physics, University of Colorado, 2000 Colorado Avenue, Boulder, CO 80309, USA}

\author[0000-0002-5463-9980]{Arvind F.\ Gupta}
\affil{Department of Astronomy \& Astrophysics, 525 Davey Laboratory, The Pennsylvania State University, University Park, PA, 16802, USA}
\affil{Center for Exoplanets and Habitable Worlds, 525 Davey Laboratory, The Pennsylvania State University, University Park, PA, 16802, USA}

\author[0000-0003-1312-9391]{Samuel Halverson}
\affil{Jet Propulsion Laboratory, California Institute of Technology, 4800 Oak Grove Drive, Pasadena, California 91109}

\author[0000-0002-1664-3102]{Fred Hearty}
\affil{Department of Astronomy \& Astrophysics, 525 Davey Laboratory, The Pennsylvania State University, University Park, PA, 16802, USA}
\affil{Center for Exoplanets and Habitable Worlds, 525 Davey Laboratory, The Pennsylvania State University, University Park, PA, 16802, USA}

\author[0000-0002-9632-9382]{Sarah E.\ Logsdon}
\affil{NSF's National Optical-Infrared Astronomy Research Laboratory, 950 N.\ Cherry Ave., 
Tucson, AZ 85719, USA}

\author[0000-0001-5000-1018]{Andrew J. Metcalf}
\affiliation{Space Vehicles Directorate, Air Force Research Laboratory, 3550 Aberdeen Ave. SE, Kirtland AFB, NM 87117, USA}
\affiliation{Time and Frequency Division, National Institute of Technology, 325 Broadway, Boulder, CO 80305, USA} 
\affiliation{Department of Physics, 390 UCB, University of Colorado Boulder, Boulder, CO 80309, USA}

\author[0000-0003-0241-8956]{Michael W.\ McElwain}
\affil{Exoplanets and Stellar Astrophysics Laboratory, NASA Goddard Space Flight Center, Greenbelt, MD 20771, USA} 

\author[0000-0002-4404-0456]{Caroline Morley}
\affil{Department of Astronomy, The University of Texas at Austin, USA}

\author[0000-0002-2488-7123]{Jayadev Rajagopal}
\affil{NSF's National Optical-Infrared Astronomy Research Laboratory, 950 N.\ Cherry Ave., 
Tucson, AZ 85719, USA}

\author[0000-0002-4289-7958]{Lawrence W. Ramsey}
\affil{Department of Astronomy \& Astrophysics, 525 Davey Laboratory, The Pennsylvania State University, University Park, PA, 16802, USA}
\affil{Center for Exoplanets and Habitable Worlds, 525 Davey Laboratory, The Pennsylvania State University, University Park, PA, 16802, USA}

\author[0000-0003-0149-9678]{Paul Robertson}
\affil{Department of Physics \& Astronomy, University of California Irvine, Irvine, CA 92697, USA}

\author[0000-0001-8127-5775]{Arpita Roy}
\affil{Space Telescope Science Institute, 3700 San Martin Dr, Baltimore, MD 21218, USA}
\affil{Department of Physics and Astronomy, Johns Hopkins University, 3400 N Charles Street, Baltimore, MD 21218, USA}

\author[0000-0002-4046-987X]{Christian Schwab}
\affil{Department of Physics and Astronomy, Macquarie University, Balaclava Road, North Ryde, NSW 2109, Australia}

\author[0000-0002-4788-8858]{Ryan C. Terrien}
\affil{Department of Physics and Astronomy, Carleton College, One North College Street, Northfield, MN 55057, USA}

\author[0000-0001-9209-1808]{John Wisniewski}
\affil{Homer L. Dodge Department of Physics and Astronomy, University of Oklahoma, 440 W. Brooks Street, Norman, OK 73019, USA}

\author[0000-0001-6160-5888]{Jason T.\ Wright}
\affil{Department of Astronomy \& Astrophysics, 525 Davey Laboratory, The Pennsylvania State University, University Park, PA, 16802, USA}
\affil{Center for Exoplanets and Habitable Worlds, 525 Davey Laboratory, The Pennsylvania State University, University Park, PA, 16802, USA}
\affil{Penn State Extraterrestrial Intelligence Center, 525 Davey Laboratory, The Pennsylvania State University, University Park, PA, 16802, USA}



\correspondingauthor{Shubham Kanodia}
\email{shbhuk@gmail.com}

\begin{abstract}
We present the discovery of a new Jovian-sized planet, TOI-3757 b, the lowest density transiting planet known to orbit an M dwarf (M0V). This planet was discovered around a solar-metallicity M dwarf, using TESS photometry and confirmed with precise radial velocities from HPF and NEID.  With a planetary radius of \plradius{} \earthradius{} and mass of \plmass{} \earthmass{}, not only does this object add to the small sample of gas giants ($\sim 10$) around M dwarfs, but also, its low density ($\rho =$  0.27$^{+0.05}_{-0.04}$ \gcmcubed{}) provides an opportunity to test theories of planet formation. We present two hypotheses to explain its low density; first, we posit that the low metallicity of its stellar host ($\sim$ 0.3 dex lower than the median metallicity of M dwarfs hosting gas giants) could have played a role in the delayed formation of a solid core massive enough to initiate runaway accretion. Second, using the eccentricity estimate of $0.14 \pm 0.06$ we determine it is also plausible for tidal heating to at least partially be responsible for inflating the radius of TOI-3757b b.  The low density and large scale height of TOI-3757 b makes it an excellent target for transmission spectroscopy studies of atmospheric escape and composition (TSM $\sim$ 190). We use HPF to perform transmission spectroscopy of TOI-3757 b using the helium 10830 \AA~ line. Doing this, we place an upper limit of 6.9 \% (with 90\% confidence) on the maximum depth of the absorption from the metastable transition of He at $\sim$ 10830 \AA, which can help constraint the atmospheric mass loss rate in this energy limited regime. 
\end{abstract}


\section{Introduction} \label{sec:intro}
Giant planets ($R_p > 4$ \earthradius{}) should be intrinsically rare around M dwarfs according to planet formation models based on the core-accretion framework \citep{laughlin_core_2004, ida_toward_2005}. In fact, recent simulations by \cite{burn_new_2021} find that gas giants do not form for host stars $< 0.5$ \solmass{}. Since the protoplanetary disk mass (and hence the amount of rocky material available) is correlated with the host star mass \citep{andrews_mass_2013}, disks around lower-mass stars such as M dwarfs should have less material available to form planetary cores. \cite{laughlin_core_2004} show that the low solid surface density of disks, coupled with the longer orbital timescales (due to lower stellar mass) makes it difficult to form these massive cores.  If the cores become massive enough \citep[$> 10$ \earthmass{}, ][]{pollack_formation_1996, ida_toward_2004_2} before the protoplanetary disk depletes, the process of runaway gas accretion is initiated, which is responsible for the formation of Jovian type planets. In addition to a potential lack of disk mass around lower-mass stars, the difficulty of forming a Jovian type planet around an M dwarf is further exacerbated by the shorter disk lifetimes in the UV-rich environment of M dwarfs, where the disks can rapidly deplete due to evaporation \citep{adams_photoevaporation_2004}.

Results from radial velocity (RV) searches \citep{endl_exploring_2006, johnson_new_2007, sabotta_carmenes_2021}, transiting \citep{kovacs_hot_2013, morton_radius_2014, obermeier_pan-planets_2016},  microlensing and direct imaging \citep{gould_frequency_2006, montet_trends_2014} studies overall support this theory by constraining the occurrence rate of short-period gas giants around M dwarfs to $\sim$ 1 -- 2\%. New and ongoing giant planet confirmations from NASA's TESS mission \citep[Transiting Exoplanet Survey Satellite;][]{ricker_transiting_2014} will help refine these estimates further, with 10 planets already added to this sample with TOI-1728b \citep{kanodia_toi-1728b_2020}, TOI-1899b \citep{canas_warm_2020}, TOI-442b \citep{dreizler_carmenes_2020}, TOI-674b \citep{murgas_toi-674b_2021}, TOI-532b \citep{kanodia_toi-532b_2021}, HATS-74Ab, HATS-75b \citep{jordan_hats-74ab_2021}, and recently TOI-3629b, and TOI-3714b \citep{canas_hot_2022}. While the sample of M dwarf gas giants is small, several general trends appear to be emerging. One, the majority of M dwarf gas giants discovered by TESS are found orbiting early-M type stars which should possess larger disks than their mid/late-M counterparts. Second, similar to the FGK orbiting hot-Jupiters \citep{gonzalez_stellar_1997, santos_metal-rich_2001, fischer_planet-metallicity_2005, ghezzi_stellar_2010, sousa_spectroscopic_2011}, there is an apparent correlation between the occurrence of gas giants and host star metallicity  \citep{johnson_metal_2009, maldonado_connecting_2019}. However, pursuing this trend further with M dwarfs is currently hampered by their intrinsic faintness (in the optical) as well as the complexities with M dwarf metallicity determination \citep{passegger_metallicities_2022}.

An important distinction between M dwarf orbiting hot-Jupiters and their FGK orbiting cousins is the large equilibrium temperature difference. As M dwarfs are significantly cooler, their hot-Jupiters possess equilibrium temperatures $<$1000 K. It is therefore unlikely that M-dwarf hot-Jupiters experience the same inflation mechanism that ``puffs'' up the radius of the hotter hot-Jupiters around FGK stars \citep[e.g. Ohmic dissipation][]{batygin.ohmic}. That said, given that most M dwarf hot-Jupiters fall within 0.05 AU, they may experience some form of tidal heating assuming they are able to maintain a slightly eccentric orbit. \citet{millholland_tidal_2020} demonstrate that even a slightly eccentric orbit could explain the inflated radii of many low-density super-puffs, suggesting that these planets do not possess unusually large H/He atmospheres but hotter than expected interiors due to tidal forces. While this study has yet to be extended to M dwarf gas giants, due to the combination of lower stellar mass and radius (and hence a higher ratio of semi-major axis to stellar radius; $a/R_*$), these planets should experience longer circularization timescales than their FGK counterparts. Therefore, it is possible that a subset of M dwarf hot-Jupiters may currently be experiencing tidal heating in their interiors.

In this manuscript we present the discovery of a low density Jovian sized planet with an inflated radius orbiting a solar-metallicity M dwarf -- TOI-3757 in the constellation of Auriga. We use a combination of photometry from TESS and ground-based instruments (RBO), high-constrast speckle imaging (NESSI), and precision RVs from the Habitable-zone Planet Finder \citep[HPF;][]{mahadevan_habitable-zone_2012, mahadevan_habitable-zone_2014} and NEID \citep[][]{halverson_comprehensive_2016, schwab_design_2016} spectrographs. We also use HPF to observe the planet during its transit and perform transmission spectroscopy to place upper limits on absorption in He 10830 \AA. In Section \ref{sec:observations} we detail these observations which are used to characterize the system. In Section \ref{sec:stellar} we describe the methodology followed to derive the stellar parameters, while Section \ref{sec:joint} details the data analysis, including the joint fitting of the photometry and RVs, and also the upper limits we place on He 10830 \AA~ absorption (Section \ref{sec:he10830}). In Section \ref{sec:discussion} we place TOI-3757 b in context of other planets around M dwarfs, and also hypothesize different mechanisms that can be responsible for its low density, while finally summarizing our findings in Section \ref{sec:conclusion}.

\begin{figure*}[!t] 
\centering
\includegraphics[width=\textwidth]{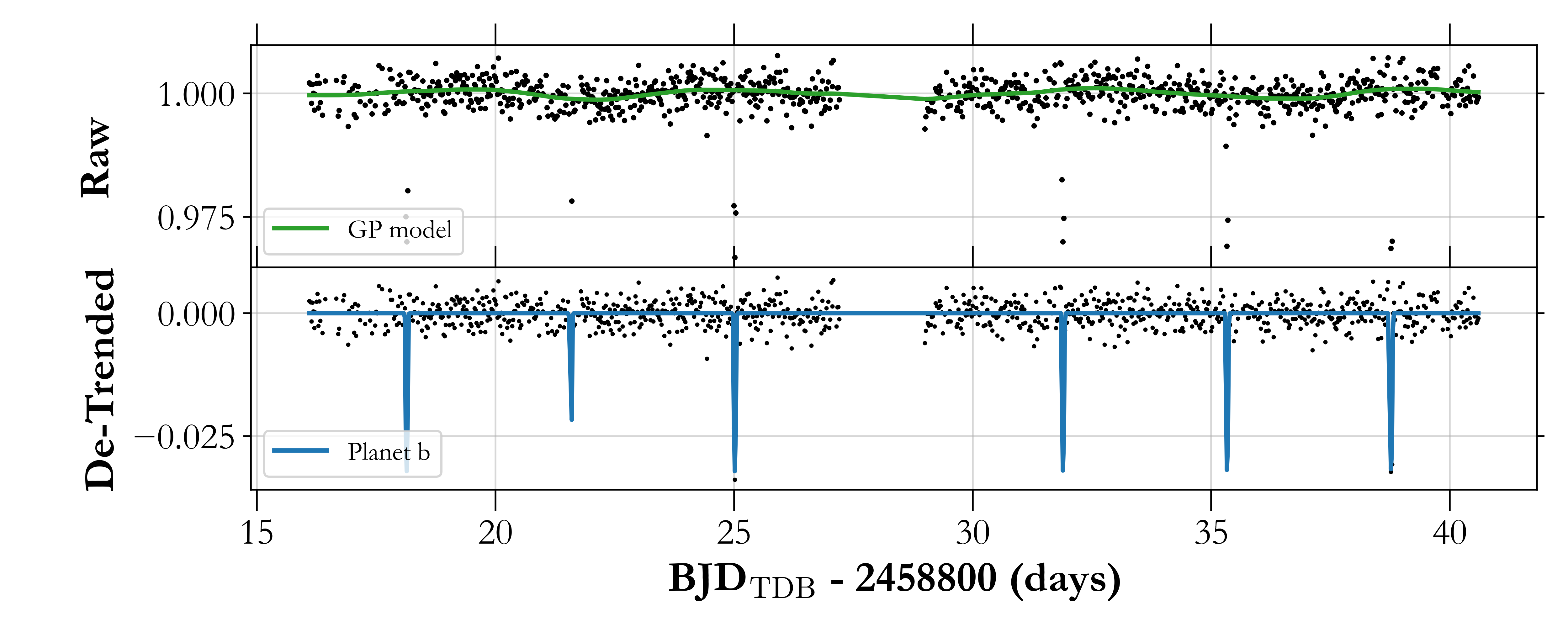}
\caption{Long cadence (30 minute) time series \tess{} \texttt{eleanor} photometry  from Sector 19, along with a stellar rotation GP kernel (\texttt{RotationTerm} from \texttt{celerite2}) in green. The detrended photometry is shown in the bottom panel, with the TOI-3757 b transits overlaid in blue.} \label{fig:tess_lc}
\end{figure*}

\section{Observations}\label{sec:observations}
\subsection{TESS}\label{sec:TESS}
TOI-3757 (TIC-445751830, 2MASS J06040089+5501126, \gaia{} EDR3 996878131494639488, UCAC4 726-038940) is an M0 dwarf observed by TESS in Sector 19 in Camera 2 from 2019 November 27 to 2019 December 24 at $\sim 30$ minute cadence (Figure \ref{fig:tess_lc}). The planet candidate was identified using the Quick Look Pipeline (QLP) algorithm developed by \cite{huang_photometry_2020}, under the `faint-star search' \citep{kunimoto_tess_2021} with a period of $\sim 3.43$ d.

We extract the photometry from the TESS full-frame images using \texttt{eleanor} \citep{feinstein_eleanor_2019}, which uses the TESScut\footnote{\url{https://mast.stsci.edu/tesscut/}} service to obtain a cut-out of \(31\times31\) pixels from the calibrated full-frame images centered on TOI-3757. The TESS light curve is derived from the \texttt{CORR\_FLUX} values, in which \texttt{eleanor} uses linear regression with pixel position, measured background, and time to remove signals correlated with these parameters. The final aperture is a $2\times1$ rectangle centered on TOI-3757 and was selected by minimizing the combined differential photometric precision (CDPP) after the data was binned in 1 hour timescales. The CDPP is formally the RMS of the photometric noise on transit timescales, and was originally defined for \textit{Kepler} \citep{jenkins_overview_2010}. We obtain a CDPP of 2730 ppm for the TESS photometry.

\begin{figure}[] 
\centering
\includegraphics[width=0.5\textwidth]{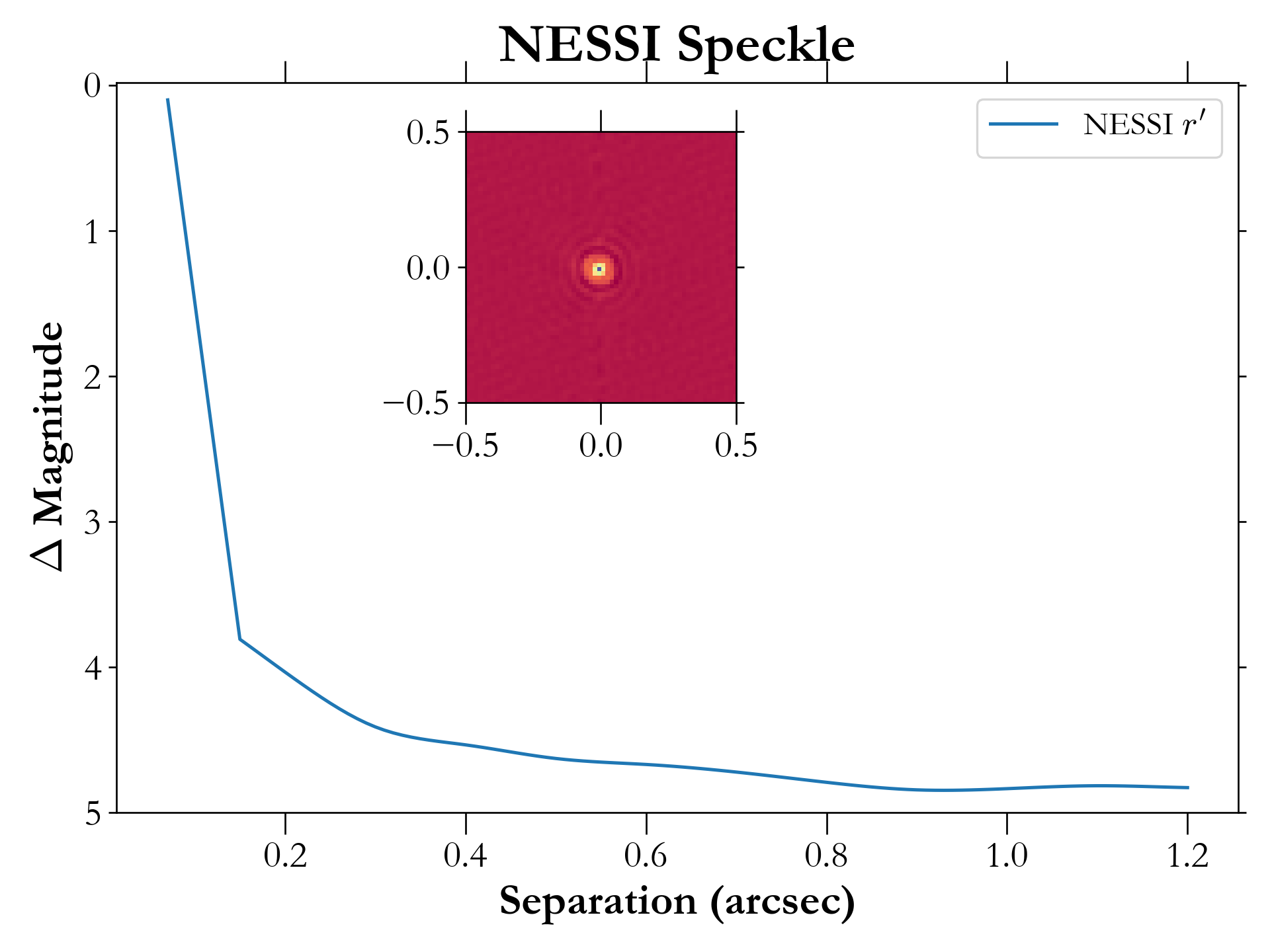}
\caption{5$\sigma$ contrast curve for TOI-3757 observed from NESSI in the Sloan \(r^\prime\)filter showing no bright companions within \(1.2''\) from the host star. The \(r^\prime\) image is shown as an inset 1$\arcsec$ across.} \label{fig:NESSI}
\end{figure}

\subsection{Ground based photometric follow up}\label{sec:photometry}

\subsubsection{RBO}
We observed a transit of TOI-3757 b on the night of 2021 November 17 using the $0.6 \unit{m}$ telescope at the Red Buttes Observatory (RBO) in Wyoming \citep{kasper_remote_2016}. The telescope is a f/8.43 Ritchey-Chrétien Cassegrain constructed by DFM Engineering, Inc, and it is currently equipped with an Apogee Alta F16 camera.

The target rose from an airmass of 1.20 at the start of the observations to a minimum airmass of 1.02 and then set to an airmass of 1.04 at the end of the observations. We performed defocussed observations using the Bessell I filter \citep{bessell_ubvri_1990} exposure times of $240 \unit{s}$. In the $2 \times 2$ binning mode, the $0.6 \unit{m}$ at RBO has a gain of $1.39 \unit{e/ADU}$, a plate scale of $0.73 \arcsec$, and a readout time of approximately $2.4 \unit{s}$. 

For the final reduction, we selected a photometric aperture of 13 pixels (9.5$\arcsec$) with a sky annulus of inner and outer radius of 23 pixels (16.8$\arcsec$) and 35 pixels (25.5$\arcsec$), respectively. We obtain an RMS precision of $\sim 1750$ ppm on the RBO photometry, after subtracting the transit model (\autoref{fig:transits}).

We observed TOI-3757 on the night of 21 December 2021 using the NN-Explore Exoplanet Stellar Speckle Imager (NESSI) on the WIYN 3.5m telescope at Kitt Peak National Observatory to search for faint background stars and nearby stellar companions. A 9-minute sequence of 40 ms diffraction-limited exposures was collected using the Sloan \(r^\prime\) filter on NESSI. The speckle images were reconstructed following the procedures described in \cite{howell_speckle_2011}. No stellar sources were detected down to a magnitude limit of $\Delta r^{\prime}$ = 4.0 at separations $>0.2''$, as shown in \autoref{fig:NESSI}.

\subsection{Radial velocity follow-up}\label{sec:rvs}

\subsubsection{HPF}\label{sec:hpf}

TOI-3757 was observed using HPF \citep{mahadevan_habitable-zone_2012, mahadevan_habitable-zone_2014} starting 2021 September 1. HPF is a near-infrared (\(8080-12780\)\ \AA), high resolution fiber-fed \citep{kanodia_overview_2018} precision RV spectrograph with exceptional environmental stability \citep{stefansson_versatile_2016} located at the 10-meter Hobby-Eberly Telescope (HET) at McDonald Observatory, Texas. HET is a fixed-altitude telescope with a roving pupil design, and is fully queue-scheduled, where all the observations are executed by the HET resident astronomers \citep{shetrone_ten_2007}. Using the algorithms described in the package \texttt{HxRGproc} \citep{ninan_habitable-zone_2018}, we correct for bias, non-linearity, cosmic rays, and calculate the slope/flux and variance images  from the raw HPF data. While HPF has the capability for simultaneous calibration using a NIR Laser Frequency Comb \citep[LFC;][]{metcalf_stellar_2019}, due to the faintness of our target we chose to avoid simultaneous calibration to minimize the impact of scattered calibrator light in the science target spectra. Instead, we obtain a wavelength solution for the target exposures by interpolating the wavelength solution from other LFC exposures on the night of the observations. This helps correct for the well-calibrated instrument drift \citep{stefansson_sub-neptune-sized_2020}.  This method has been shown to enable precise wavelength calibration and drift correction with a precision of $\sim30$ \cms{} per observation, a value much smaller than our estimated per observation RV uncertainty (instrumental + photon noise) for this object of 34 \ms{} (in 969 s exposures, and 23 \ms{} in binned 30 minute exposures). 

\begin{deluxetable}{cccc}
\tablecaption{RVs (binned in $\sim$ 30 minute exposures) of TOI-3757.  We include this table in a machine readable format along with the manuscript.\label{tab:rvs}}
\tablehead{\colhead{$\unit{BJD_{TDB}}$}  &  \colhead{RV}   & \colhead{$\sigma$} & \colhead{Instrument} \\
           \colhead{(d)}   &  \colhead{\ms{}} & \colhead{\ms{}} & \colhead{}}
\startdata
2459458.98300 & -28.29 & 22.93 & HPF \\ 
2459466.97200 & 42.47 & 24.02 & HPF \\ 
2459472.95600 & -11.67 & 23.40 & HPF \\ 
2459489.92300 & -62.93 & 21.99 & HPF \\ 
2459503.87500 & -36.31 & 28.99 & HPF \\ 
2459506.86400 & -71.04 & 23.27 & HPF \\ 
2459507.87000 & -22.90 & 20.91 & HPF \\ 
2459509.85300 & -50.03 & 23.73 & HPF \\ 
2459512.83800 & -75.03 & 22.47 & HPF \\ 
2459513.83100 & -109.27 & 28.70 & HPF \\ 
2459516.84000 & -96.78 & 18.73 & HPF \\ 
2459517.83600 & -52.47 & 19.65 & HPF \\ 
2459532.80200 & -2.53 & 26.69 & HPF \\ 
2459533.78400 & -67.68 & 28.71 & HPF \\ 
2459551.75800 & -139.99 & 25.00 & HPF \\ 
2459571.69600 & -102.35 & 25.02 & HPF \\ \hline
2459505.96900 & -0.50 & 17.21 & NEID \\ 
2459523.93279 & -36.64 & 6.78 & NEID \\ 
2459528.80595 & 66.52 & 10.79 & NEID \\ 
2459531.81514 & 27.19 & 11.85 & NEID \\ 
2459532.81927 & 46.97 & 10.55 & NEID \\ 
2459534.03339 & -43.36 & 11.05 & NEID \\ 
2459546.78516 & 55.86 & 12.63 & NEID \\ 
2459554.99677 & -21.28 & 7.57 & NEID \\ 
2459582.88049 & 19.91 & 12.05 & NEID \\ 
2459585.89929 & -20.71 & 8.26 & NEID \\ 
2459589.89730 & -1.91 & 11.30 & NEID \\ 
\enddata
\end{deluxetable}

We follow the method described in \cite{stefansson_sub-neptune-sized_2020} to derive the RVs, by using a modified version  of the \texttt{SpEctrum Radial Velocity AnaLyser} pipeline \citep[\texttt{SERVAL};][]{zechmeister_spectrum_2018}. \texttt{SERVAL} uses the template-matching method  \citep[e.g.,][]{anglada-escude_harps-terra_2012}, where it creates a master template from the target star observations, and determines the Doppler shift for each individual observation by minimizing the \(\chi^2\) statistic. The master template is created using all of the HPF observations for TOI-3757, after masking out the telluric and sky-emission lines. The telluric regions are identified by a synthetic telluric-line mask generated from \texttt{telfit} \citep{gullikson_correcting_2014}, a Python wrapper to the Line-by-Line Radiative Transfer Model package \citep{clough_atmospheric_2005}. We use \texttt{barycorrpy} to perform the barycentric correction on the individual spectra, which is the Python implementation \citep{kanodia_python_2018} of the algorithms from \cite{wright_barycentric_2014}. 

We obtained a total of 25 visits on this target between 2021 September 1 and 2021 December 24, of which 9 had to be excluded from further analysis due to poor weather conditions during the observations (seeing and sky transparency)\footnote{The rejected visits had an unbinned median S/N and RV uncertainty of $\sim$ 32 and $\sim 53$ \ms{}, compared to 47 and 34 \ms{} respectively, for the visits included in the analysis.}. Each visit was divided into 2 exposures of 969 s each, where the median S/N of each HPF exposure was 47 per pixel at 1070 nm. The individual exposures were then binned after weighting based on SNR, with the final binned RVs being listed in  \autoref{tab:rvs}.

Additionally, we observed TOI-3757 with HPF on 2021 December 12, during the transit of planet b. This observation consisted of 10x individual exposures of 649 s each, and was used to place constraints on atmospheric escape using the infrared atomic transitions of helium as a tracer (discussed further in Section \ref{sec:he10830}). Out of an abundance of caution, we do not include this visit in our RV analysis, to avoid potential systematics from the Rossiter-McLaughlin effect \citep{rossiter_detection_1924, mclaughlin_results_1924, triaud_rossiter-mclaughlin_2018}.

\subsubsection{NEID}\label{sec:neid}

We also observed TOI-3757 using NEID, a new ultra-precise \citep{halverson_comprehensive_2016}, environmentally stabilized \citep{robertson_ultrastable_2019} spectrograph at the WIYN 3.5 m telescope at Kitt Peak National Observatory. NEID is a high-resolution spectrograph ($R \sim 110,000$) with an extended red wavelength coverage \citep[380 -- 930 nm;][]{schwab_design_2016}     ; it has a fiber-feed system similar to HPF \citep{kanodia_overview_2018} with three fibers --- science, sky, and simultaneous calibration. For these observations we use the NEID high resolution (HR) mode\footnote{Instead of the high efficiency (HE) mode with resolution $R \sim 70,000$.} with resolution $R \sim 110,000$. Between 2021 November 1 and 2022 January 10, we obtained 12 visits on TOI-3757 with NEID, of which we exclude the visit from 2021 November 2 with a S/N of 2.3 at 850 nm due to patchy clouds during the observation. Of the visits included for analysis, each consisted of an 1800 s exposure with a median S/N per 1D extracted pixel of 9.3 at 850 nm. 

The NEID data was reduced using the NEID Data Reduction Pipeline\footnote{\url{https://neid.ipac.caltech.edu/docs/NEID-DRP/}} (DRP), and the Level-2 1D extracted spectra were retrieved from the NEID Archive\footnote{\url{https://neid.ipac.caltech.edu/}}. We used a modified version of the  \texttt{SERVAL} template-matching algorithm to obtain the NEID RVs \citep{stefansson_warm_2021}, similar to that for HPF. The NEID RVs presented here were calculated using orders spanning 4560 -- 8960 \AA~(order indices 40 to 104), across the central 7000 pixels which masks out the low S/N spectra outside the free spectral range. We also mask out the telluric and sky emission lines which were identified following an identical procedure to our HPF processing (Section \ref{sec:hpf}), and obtain the barycentric velocities using \cite{kanodia_python_2018}. The final NEID RVs are listed in \autoref{tab:rvs}.

\begin{figure*}[!t] 
\centering
\includegraphics[width=\textwidth]{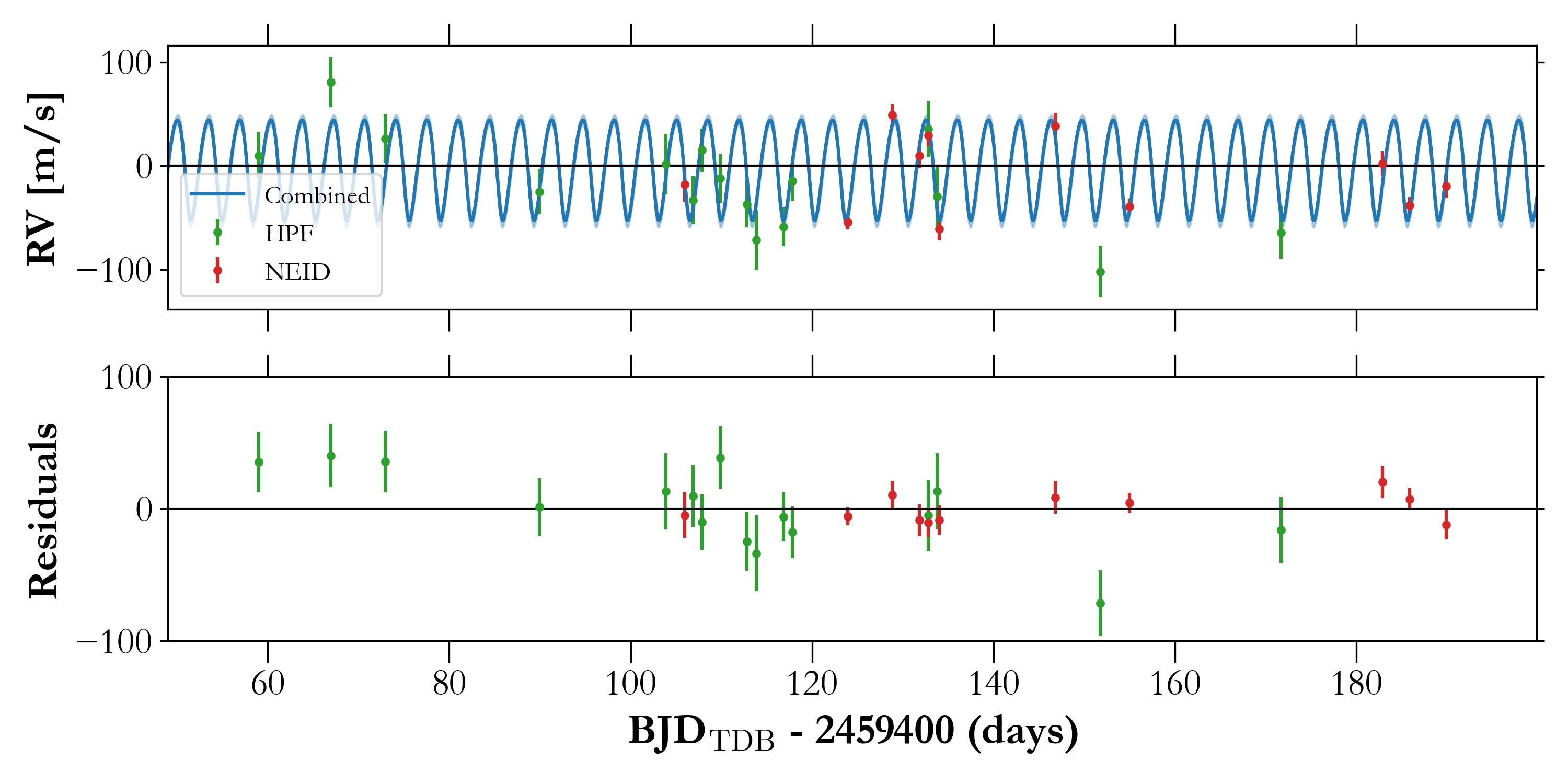}
\caption{Time series of RV observations of TOI-3757 with HPF (green) and NEID (red). The best-fitting model derived from the joint fit to the photometry and RVs is plotted in blue, including the 16-84$\%$ confidence interval in lighter blue. The bottom panel shows the residuals after subtracting the model.} \label{fig:rv}
\end{figure*}

The early spectral type (M0) of TOI-3757 leads to higher RV information in the optical, than in the NIR \citep{bouchy_fundamental_2001}. Coupled with negligible rotational broadening (\vsini{} $< 2$ \kms{}), this leads to NEID RVs ($R \sim 110,000$) that are much more precise than HPF ($R \sim 55,000$; \autoref{tab:rvs}, \autoref{fig:rv}).

\section{Stellar Parameters}\label{sec:stellar}

\subsection{Spectroscopic Parameters with HPF-SpecMatch}

We use the \texttt{HPF-SpecMatch}\footnote{\url{https://gummiks.github.io/hpfspecmatch/}} package \citep{stefansson_sub-neptune-sized_2020} to empirically determine stellar parameters from HPF spectra using the template matching method based on \cite{yee_precision_2017}.

The \texttt{HPF-SpecMatch} algorithm employs a two-step process, which uses the $\chi^2$ metric twice to find the library stars that best fit the target spectrum. The HPF spectral library contains 166 stars and spans the following parameter space: $2700~\mathrm{K} < T_{e} < 6000~\mathrm{K}$, $4.3<\log g_\star < 5.3$, and $-0.5 < \mathrm{[Fe/H]} < 0.5$. 

In the first step, each stellar library spectrum is compared with the target spectrum using the $\chi^2$ metric and ranked from best to worst-fitting library star to the target spectrum. In the second step, only the top 5 best-fit ranked library stars are used. The $\chi^2$ metric is applied to assign scaling constants to each of the 5 best-fit library stars and create a composite spectrum that fits the target spectrum even more closely. These scaling constants are also used to determine a weighted average for the precise parameter estimates of effective temperature \teff{}, surface gravity (log $g$), and metallicity ([Fe/H]) of the target star.

The errors we adopt for the spectroscopic parameters are determined via a leave-one-out cross-validation, broadly following the methodology in \cite{stefansson_sub-neptune-sized_2020}. In this process, a library star of interest is removed from the rest of the stellar library pool. Then, the \texttt{HPF-SpecMatch} algorithm is run to estimate its stellar parameters independently of its true parameter. The difference between these calculated parameters and the true parameter values are noted. The spectral matching is performed on HPF order index 5 ($8534-8645$ \AA) for TOI-3757 because this order has negligible telluric contamination. The resolution limit of HPF places a constraint of $v \sin i < 2 \mathrm{~km~s^{-1}}$ for TOI-3757. Table \ref{tab:stellarparam} presents the derived spectroscopic parameters with their uncertainties.

\subsection{Model-Dependent Stellar Parameters}\label{sec:stellarparams}
We derive model-dependent stellar parameters by modeling the spectral energy distribution (SED) for TOI-3757 using the {\tt EXOFASTv2} analysis package \citep{eastman_exofast_2013}. The SED fit uses the precomputed bolometric corrections\footnote{\url{http://waps.cfa.harvard.edu/MIST/model_grids.html\#bolometric}} in \(\log g_\star\), \teff{}, [Fe/H], and \(A_V\) from the MIST model grids \citep{dotter_mesa_2016, choi_mesa_2016}. 

We place Gaussian priors on the (i) broadband photometry listed in Table \ref{tab:stellarparam}; (ii) the spectroscopic stellar parameters derived with \texttt{HPF-SpecMatch}, and (iii) the geometric distance calculated from \cite{bailer-jones_estimating_2021}. We set the upper limit of the visual extinction to the estimate from \cite{green_3d_2019} calculated at the distance determined by \cite{bailer-jones_estimating_2021}. The extinction from \cite{green_3d_2019} is converted to a visual magnitude extinction using the \(R_{v}=3.1\) reddening law from \cite{fitzpatrick_correcting_1999}. \autoref{tab:stellarparam} contains the stellar priors and derived stellar parameters with their uncertainties. 

\subsection{Galactic kinematics}

We use the systemic velocity derived from HPF and proper motion from GAIA EDR3 to calculate the \textit{UVW} velocities in the barycentric frame using \texttt{GALPY} \citep{bovy_galpy_2015}\footnote{With \textit{U} towards the Galactic center, \textit{V} towards the direction of Galactic spin, and \textit{W} towards the North Galactic Pole \citep{johnson_calculating_1987}.}. We also provide the UVW velocities in the local standard of rest using the offsets from \cite{schonrich_local_2010}. Using the BANYAN tool \citep{gagne_banyan_2018}, we classify TOI-3757 as a field star in the thin disk with very high probability \citep[$> 99 \%$;][]{bensby_exploring_2014}.


\subsection{Rotation Period Estimates}\label{sec:rotation}

\begin{figure}[] 
\centering
\includegraphics[width=0.5\textwidth]{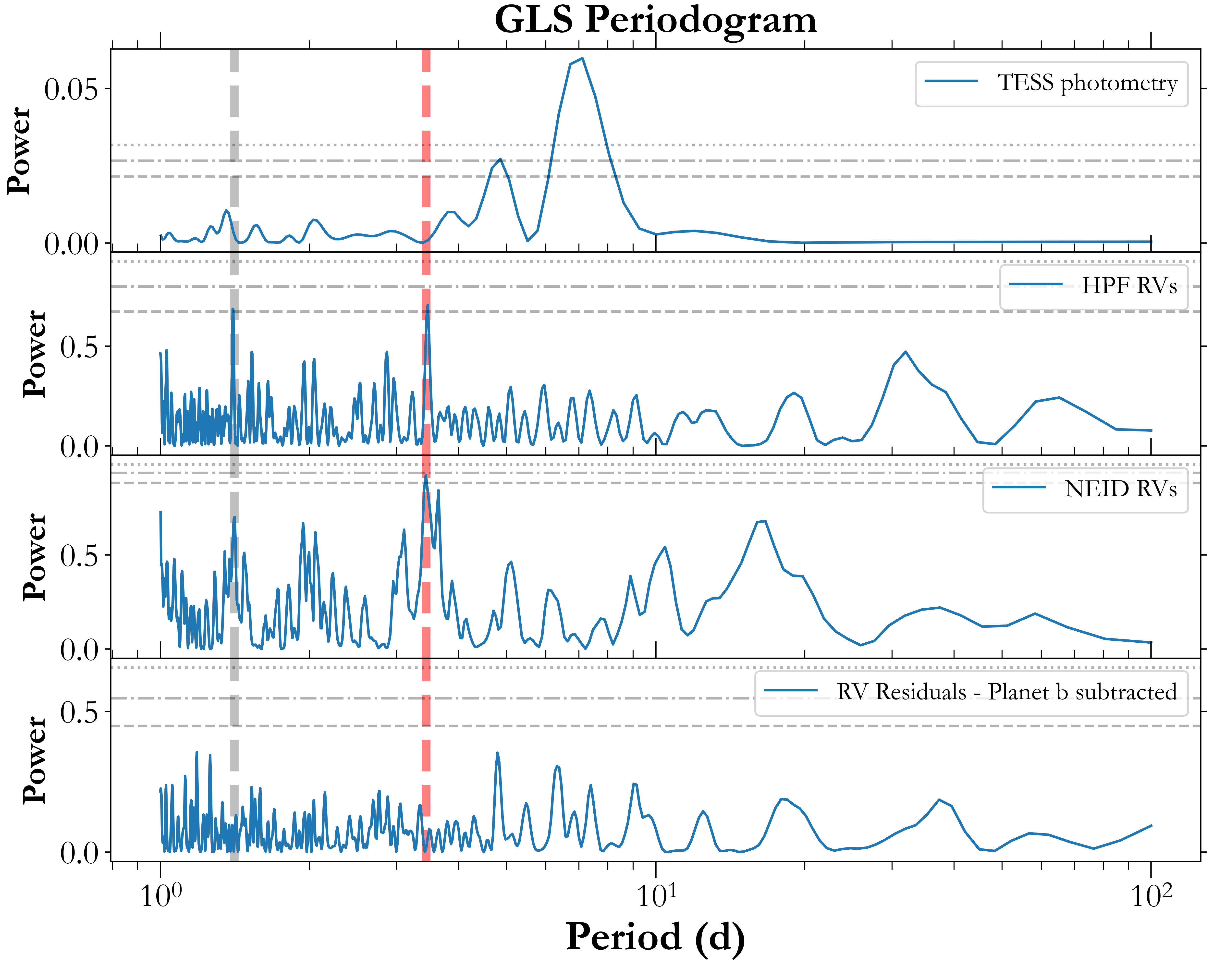}
\caption{GLS periodogram for the different datasets. The horizontal grey lines represent the 0.1\%, 1\%, and 10\% FAP values. The red vertical line depicts the orbital period of planet b, whereas the grey line marks its 1-day alias.} \label{fig:GLS}
\end{figure}

We  run a generalized Lomb Scargle (GLS) periodogram \citep{lomb_least-squares_1976,scargle_studies_1982, zechmeister_generalised_2009} on the \tess{} photometry (after masking the transits on TOI-3757 b) using its \texttt{astropy} implementation, and find a significant peak ($< 0.1 \%$ False Alarm Probability) at $\sim 7$ days (\autoref{fig:GLS}). This is consistent with the results from our Gaussian Process (GP)  stellar rotation kernel\footnote{\texttt{RotationTerm} implemented in \texttt{celerite2} \citep{foreman-mackey_fast_2017, foreman-mackey_scalable_2018}.} applied to the \tess{} photometry, which suggest a stellar rotation period of $6.9^{+0.5}_{-0.7}$ days. This kernel consists of two simple harmonic oscillator terms -- one at the rotation period, with the second one at half the period.

On further inspection using \texttt{eleanor} we find a similar significant peak at $\sim 7$ days for the photometry in the majority of the adjoining pixels in a 6x6 grid centered on the centroid for TOI-3757. This suggests that the periodic signal seen in the \texttt{eleanor} reduction of the \tess~ FFI photometry is not astrophysical in origin, and TOI-3757 does not have significant rotational modulation. This is further corroborated by the lack of a detectable rotational broadening signal in the HPF spectra, using which we can place a limit of  \vsini{} $< 2$ \kms{} on the host star\footnote{The corresponding equatorial velocity for a $\sim 7$ day rotation period would be $\sim 4.5$ \kms{}.}.

We also access publicly available data from the Zwicky Transient Facility \citep[ZTF,][]{masci_zwicky_2019} and ASAS-SN \citep{kochanek_all-sky_2017} for this target to perform a GLS periodogram analysis, and do not detect any rotation signal present in the photometry. The photometry spans $\sim$ 800 days for ZTF and $\sim$ 1000 days for ASAS-SN.  We also analyze the Ca infrared triplet from the HPF spectra, and H$\alpha$ from the NEID spectra, but do not find any periodic signals in the time series.

\begin{deluxetable*}{lccc}
{\tabletypesize{\footnotesize }
\tablecaption{Summary of stellar parameters for TOI-3757. \label{tab:stellarparam}}
\tablehead{\colhead{~~~Parameter}&  \colhead{Description}&
\colhead{Value}&
\colhead{Reference}}
\startdata
\multicolumn{4}{l}{\hspace{-0.2cm} Main identifiers:}  \\
~~~TOI & \tess{} Object of Interest & 3757 & \tess{} mission \\
~~~TIC & \tess{} Input Catalogue  & 445751830 & Stassun \\
~~~2MASS & \(\cdots\) & J06040089+5501126 & 2MASS  \\
~~~Gaia EDR3 & \(\cdots\) & 996878131494639488 & Gaia EDR3\\
\multicolumn{4}{l}{\hspace{-0.2cm} Equatorial Coordinates, Proper Motion and Spectral Type:} \\
~~~$\alpha_{\mathrm{J2016}}$ &  Right Ascension (RA) & 06:04:00.87 & Gaia EDR3\\
~~~$\delta_{\mathrm{J2016}}$ &  Declination (Dec) & 55:01:11.90 & Gaia EDR3\\
~~~$\mu_{\alpha}$ &  Proper motion (RA, \unit{mas/yr}) &  $-9.03 \pm 0.02$ & Gaia EDR3\\
~~~$\mu_{\delta}$ &  Proper motion (Dec, \unit{mas/yr}) & $-43.13 \pm 0.02$ & Gaia EDR3 \\
~~~$d$ &  Distance in pc  & $177.4 \pm 0.7$ & Bailer-Jones \\
~~~\(A_{V,max}\) & Maximum visual extinction & 0.26 & Green\\
\multicolumn{4}{l}{\hspace{-0.2cm} Optical and near-infrared magnitudes:}  \\
~~~$B$ & Johnson B mag & $16.2 \pm 0.2$ & APASS\\
~~~$V$ & Johnson V mag & $14.8 \pm 0.1$ & APASS\\
~~~$g^{\prime}$ &  Sloan $g^{\prime}$ mag  & $15.5 \pm 0.1$ & APASS\\
~~~$r^{\prime}$ &  Sloan $r^{\prime}$ mag  & $14.2 \pm 0.1$ & APASS \\
~~~$i^{\prime}$ &  Sloan $i^{\prime}$ mag  & $13.5 \pm 0.1$ & APASS \\
~~~$J$ & $J$ mag & $12.00 \pm 0.03$ & 2MASS\\
~~~$H$ & $H$ mag & $11.31 \pm 0.03$ & 2MASS\\
~~~$K_s$ & $K_s$ mag & $11.15 \pm 0.02$ & 2MASS\\
~~~$W1$ &  WISE1 mag & $11.06 \pm 0.02$ & WISE\\
~~~$W2$ &  WISE2 mag & $11.10 \pm 0.02$ & WISE\\
~~~$W3$ &  WISE3 mag & $11.0 \pm 0.1$ & WISE\\
\multicolumn{4}{l}{\hspace{-0.2cm} Spectroscopic Parameters$^a$:}\\
~~~$T_{\mathrm{eff}}$ &  Effective temperature in \unit{K} & $3913 \pm 56$ & This work\\
~~~$\mathrm{[Fe/H]}$ &  Metallicity in dex & $0.0 \pm 0.20$ & This work\\
~~~$\log(g)$ & Surface gravity in cgs units & $4.68 \pm 0.04$ & This work\\
\multicolumn{4}{l}{\hspace{-0.2cm} Model-Dependent Stellar SED and Isochrone fit Parameters$^b$:}\\
~~~$M_*$ &  Mass in $M_{\odot}$ & $0.64 \pm 0.02$ & This work\\
~~~$R_*$ &  Radius in $R_{\odot}$ & $0.62 \pm 0.01$ & This work\\
~~~$L_*$ &  Luminosity in $L_{\odot}$ & $0.087 \pm 0.003$ & This work\\
~~~$\rho_*$ &  Density in \gcmcubed{} & $3.7 \pm 0.2$ & This work\\
~~~Age & Age in Gyrs & $7.1\pm4.5$ & This work\\
~~~$A_v$ & Visual extinction in mag & $0.067^{+0.078}_{-0.047}$ & This work\\
\multicolumn{4}{l}{\hspace{-0.2cm} Other Stellar Parameters:}           \\
~~~$v \sin i_*$ &  Rotational velocity in \kms{}  & $< 2$ \kms{} & This work\\
~~~$\Delta$ RV &  ``Absolute'' radial velocity in \kms{} & $21.86\pm0.04$ & This work\\
~~~$U, V, W$ &  Galactic velocities in \kms{} &  $-36.12\pm0.07, -17.81\pm0.10, -15.82\pm0.09$ & This work\\
~~~$U, V, W^c$ &  Galactic velocities (LSR) in \kms{} & $-25.02\pm0.85, -5.57\pm0.69, -8.56\pm0.61$ & This work\\
\enddata
\tablenotetext{}{References are: Stassun \citep{stassun_tess_2018}, 2MASS \citep{cutri_2mass_2003}, Gaia EDR3 \citep{collaboration_gaia_2021}, Bailer-Jones \citep{bailer-jones_estimating_2018}, Green \citep{green_3d_2019}, APASS \citep{henden_apass_2018}, WISE \citep{wright_wide-field_2010}}
\tablenotetext{a}{Derived using the HPF spectral matching algorithm from \cite{stefansson_sub-neptune-sized_2020}}
\tablenotetext{b}{{\tt EXOFASTv2} derived values using MIST isochrones with the \gaia{} parallax and spectroscopic parameters in $a$) as priors.}
\tablenotetext{c}{The barycentric UVW velocities are converted into local standard of rest (LSR) velocities using the constants from \cite{schonrich_local_2010}.}
}
\end{deluxetable*}

\begin{figure*}[] 
\centering
\includegraphics[width=\textwidth]
{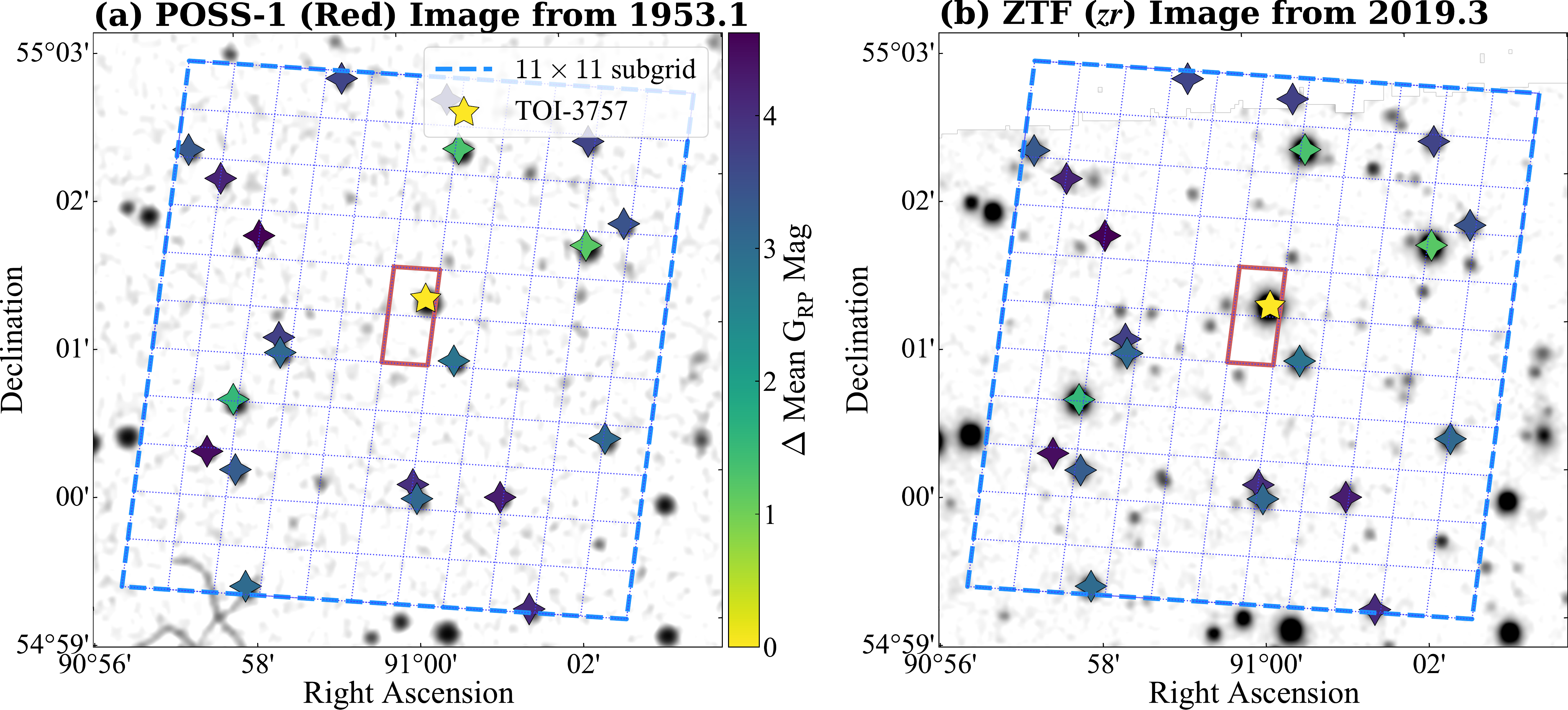}
\caption{\textbf{Panel a} overlays an 11 x 11 pixel footprint from \tess{} Sector 19 (blue grid) on a POSS-I red image from 1953.1. TOI-3757 has a small proper motion as can be seen while comparing Panel \textbf{a)} and \textbf{b)}. The \tess{} aperture is outlined in red and we highlight TOI-3757 with a star. No bright targets are present inside the \tess{} aperture with $\Delta$ G$_{RP}$ $<$ 3. \textbf{Panel b} is similar to Panel A but with a background image from ZTF \(zr\) (5600 \AA -- 7316 \AA) from 2019 \citep{masci_zwicky_2019}.} \label{fig:tess_map}
\end{figure*}

\begin{figure*}[] 
\centering
\includegraphics[width=1.1\textwidth]{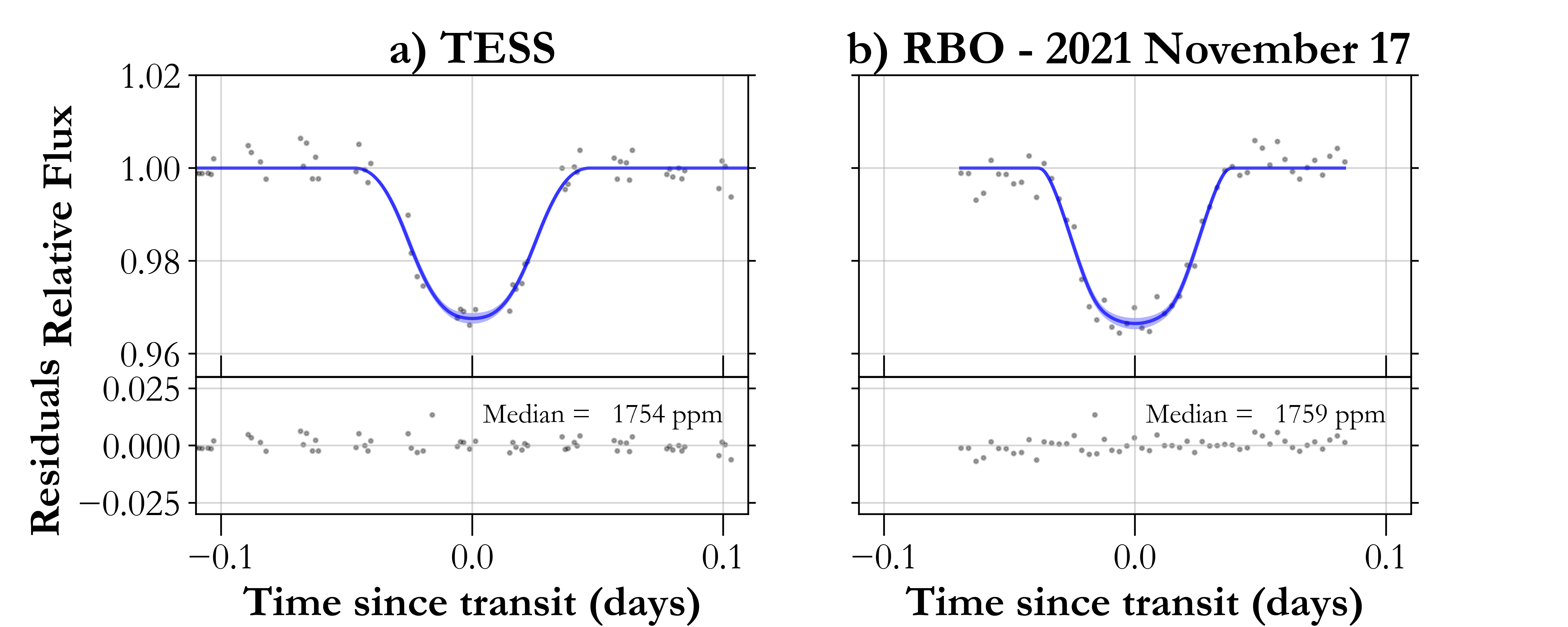}
\caption{Photometric observations for TOI-3757 b; a) the TESS phased plot shows the detrended light curve phase-folded to the best fit orbital period, b) Ground based observations for TOI-3757 b from RBO. The raw photometery is shown in grey, with the best-fit transit solution in blue, along with the 1 $\sigma$ confidence interval shown in lighter-blue. } \label{fig:transits}
\end{figure*}

\subsection{Ruling out Stellar Companions}\label{sec:stellar companions}
\subsubsection{Unresolved Stellar Companions}
We try to place constraints on the presence of unresolved stellar comapnions using HPF spectra, Gaia astrometry and the RVs:
\begin{itemize}
    \item Constraints from HPF spectra: We follow the procedure outlined in \cite{kanodia_toi-1728b_2020}, to place limits on any spatially unresolved stellar companion to TOI-3757 using the HPF spectra to quantify the lack of flux from a secondary object. We combine the spectra from a single epoch to obtain a higher S/N template for comparison, and then model the test spectra (TOI-3757) as a linear combination of a primary M dwarf (BD+29\_2279) and a secondary companions (GJ 251, GJ 1156 and VB-10). The flux ratio between the secondary and primary star, $F$, is calculated as:
        
    \begin{eqnarray}
    S_{\mathrm{obs}} &=& A \left( (1-x)S_{\mathrm{primary}} + (x)S_{\mathrm{secondary}} \right) \label{eq:spectra} \\
    F &=& \frac{x}{1-x} \label{eq:fluxratio}
    \end{eqnarray}
    \noindent where $S_{\mathrm{obs}}$ is the observed spectrum,  $S_{\mathrm{primary}}$ is the primary spectrum, $S_{\mathrm{secondary}}$ represents the secondary spectrum, and $A$ is the normalization constant. For a given primary and secondary template, we (i) perform a $\chi^2$ minimization to shift the secondary spectrum in velocity space, (ii) add this shifted secondary spectrum to the primary, and (iii) fit for the value of $x$ (and $A$) that best fits the observed spectrum. We perform this for a range of spectral types for the secondary from M3 to M8 spanning velocity offsets of $\pm 150$ \kms{}. We place a conservative upper limit for a secondary companion of flux ratio $<$ 0.15 or $\Delta \rm{mag} \simeq 2.1$ for $|\Delta v|$  $> $ 5 \kms{}, using HPF order index 17 spanning $10450 - 10580$ \AA. The lower limit coincides with HPF's spectral resolution ($R \sim 55,000 \approx 5.5$~\kms{}). At lower velocity offsets, the degeneracy between the primary and secondary spectra prevents any meaningful flux ratio constraints
    
    \item Constraints from Gaia astrometry: GAIA EDR3 \citep{lindegren_gaia_2020} provides an additional astrometric constraint on the presence of unresolved bound companions using the re-normalized unit weight error (RUWE) metric. RUWE is sensitive to the change in the position of the primary target due to reflex motion caused by unresolved bound companions. For the single-star astrometric solution in use for GAIA EDR3, this astrometric motion of the primary star around the center of mass would manifest as noise \citep{kervella_stellar_2019}, especially for orbital periods much shorter than the observing baseline for GAIA EDR3 ($\sim 34$ months). The commonly accepted threshold in literature for this is RUWE $\gtrsim 1.4$, which correlates with the presence of a bound stellar companion in recent studies of stellar binaries \citep{penoyre_binary_2020, belokurov_unresolved_2020, gandhi_astrometric_2021}. For TOI-3757, GAIA EDR3 reports a RUWE of $\sim 1.1$, which is in agreement with a single-star astrometric solution.

    \item Constraints from RVs: We use the joint fit of the photometry and radial velocity to estimate the planetary and system properties (Section \ref{sec:joint}). We also include a linear RV trend in the orbital solution while fitting the RVs, and note this to be consistent with 0, where the RV trend $\sim$ -0.23$^{+4.76}_{-4.89}$ (\ms{} yr$^{-1}$). The residuals to this fit (Shown in \autoref{fig:rv}), are also analyzed with a GLS periodogram and show no significant signals (\autoref{fig:GLS}), indicating the absence of any long-period bound companions over our observing baselines ($\sim 100$ days).

\end{itemize}

\subsubsection{Resolved Stellar Companions}
\autoref{fig:tess_map} presents a comparison of the region contained in the $11\times11$ pixel footprint from Sector 19 using a Palomar Observatory Sky Survey \citep[POSS-1;][]{harrington_48-inch_1952, minkowski_national_1963} image from 1951 and a ZTF \citep{masci_zwicky_2019} image from 2019. The POSS-1 plate images were taken with Eastman 103a-E spectroscopic plates in conjunction with a No. 160 red plexiglass filter with a bandpass between 6000 -- 6700 \AA, and have a limiting magnitude of $R \sim 19$ \citep{harrington_48-inch_1952}. TOI-3757 has low proper-motion, and the change in coordinates between the two epochs is negligible. There are no bright targets with $\Delta$ G$_{RP}$ $<$ 3 present in the $2\times1$ \tess{} aperture. There are a few targets with $\Delta$~G$_{RP}$ $<$ 4, that may dilute the \tess{} transit. We use the ground based photometry to estimate the dilution term for the TESS photometry. Additionally, using the NESSI observations we are able to rule out stellar sources with a $\Delta r^{\prime} < 4.0$ at separations $>0.2''$.

\begin{figure}[] 
\centering
\includegraphics[width=0.5\textwidth]{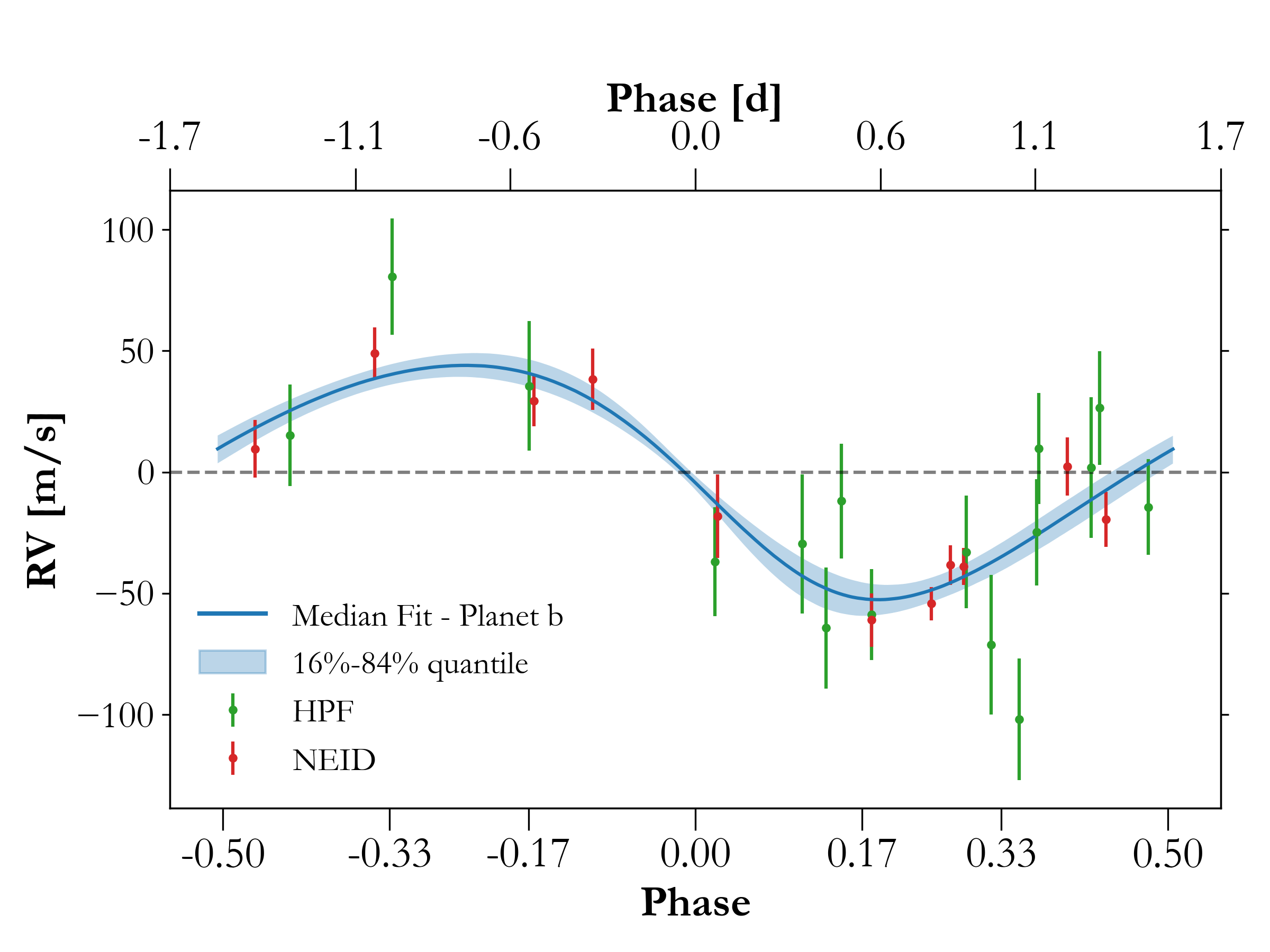}
\caption{HPF + NEID RV observations phase folded on the best fit orbital period from the joint fit from Section \ref{sec:joint}. The best fit model is shown in the solid line, whereas the $1\sigma$ confidence intervals are shown in lighter blue.} \label{fig:RVphase}
\end{figure}

\section{Data Analysis}
\subsection{Joint Fitting of Photometry and RVs}\label{sec:joint}

\begin{deluxetable*}{llc}
\tablecaption{Derived Parameters for the TOI-3757 System.  \label{tab:planetprop}}
\tablehead{\colhead{~~~Parameter} &
\colhead{Units} &
\colhead{Value$^a$} 
}
\startdata
\sidehead{Orbital Parameters:}
~~~Orbital Period\dotfill & $P$ (days) \dotfill & 3.438753$\pm$0.000004\\
~~~Eccentricity\dotfill & $e$ \dotfill & 0.14$\pm0.06$ \\
~~~Argument of Periastron\dotfill & $\omega$ (degrees) \dotfill & 130$\pm23$ \\
~~~Semi-amplitude Velocity\dotfill & $K$ (\ms{})\dotfill &
49.24$\pm5.07$\\
~~~Systemic Velocity$^b$\dotfill & $\gamma_{\mathrm{HPF}}$ (\ms{})\dotfill & -37.91$^{+7.80}_{-7.85}$\\
~~~ & $\gamma_{\mathrm{NEID}}$ (\ms{})\dotfill & 17.58$\pm4.15$\\
~~~RV trend\dotfill & $dv/dt$ (\ms{} yr$^{-1}$)   & -0.20$^{+4.96}_{-4.94}$   \\ 
~~~RV jitter\dotfill & $\sigma_{\mathrm{HPF}}$ (\ms{})\dotfill & 17.71$^{+10.01}_{-9.49}$\\
~~~ & $\sigma_{\mathrm{NEID}}$ (\ms{})\dotfill & 5.82$^{+5.66}_{-3.99}$\\
\sidehead{Transit Parameters:}
~~~Transit Midpoint \dotfill & $T_C$ (BJD\textsubscript{TDB})\dotfill & 2458838.77148$^{+0.00062}_{-0.00061}$\\
~~~Scaled Radius\dotfill & $R_{p}/R_{*}$ \dotfill & 
0.1769$^{+0.0056}_{-0.0065}$\\
~~~Scaled Semi-major Axis\dotfill & $a/R_{*}$ \dotfill & 13.26$^{+0.26}_{-0.25}$\\
~~~Orbital Inclination\dotfill & $i$ (degrees)\dotfill & 86.76$^{+0.23}_{-0.20}$\\
~~~Transit Duration\dotfill & $T_{14}$ (days)\dotfill & 0.0800$^{+0.0038}_{-0.0030}$\\
~~~Photometric Jitter$^c$ \dotfill & $\sigma_{TESS}$ (ppm)\dotfill & $2569_{-59}^{+62}$\\ 
~~~ & $\sigma_{\mathrm{RBO20211117}}$ (ppm)\dotfill & $2953_{-283}^{+332}$\\ 
~~~Limb Darkening$^d$ $\dotfill$ & $u_{1,\rm{TESS}}$, $u_{2,\rm{TESS}}$ $\dotfill$ & $0.39^{+0.41}_{-0.28}$, $0.10^{+0.38}_{-0.33}$  \\
 & $u_{1,\rm{RBO20211117}}$, $u_{2,\rm{RBO20211117}}$ $\dotfill$ & $0.59^{+0.43}_{-0.40}$, $0.07^{+0.43}_{-0.41}$ \\
~~~Dilution$^e$\dotfill & $D_{\mathrm{TESS}}$ \dotfill & 0.994$\pm0.053$\\
\sidehead{Planetary Parameters:}
~~~Mass\dotfill & $M_{p}$ (M$_\oplus$)\dotfill &  \plmass{}\\
~~~Radius\dotfill & $R_{p}$  (R$_\oplus$) \dotfill& \plradius{}\\
~~~Density\dotfill & $\rho_{p}$ (\gcmcubed{})\dotfill & 0.27$^{+0.05}_{-0.04}$\\
~~~Semi-major Axis\dotfill & $a$ (AU) \dotfill & 0.03845$\pm0.00043$\\
~~~Average Incident Flux$^f$\dotfill & $\langle F \rangle$ (\unit{10^5\ W/m^2})\dotfill &  0.75$\pm$0.05\\
~~~Planetary Insolation& $S$ (S$_\oplus$)\dotfill &  55.4$\pm3.8$\\
~~~Equilibrium Temperature$^g$ \dotfill & $T_{\mathrm{eq}}$ (K)\dotfill & 759$\pm13$\\
\enddata
\tablenotetext{a}{The reported values refer to the 16-50-84\% percentile of the posteriors.}
\tablenotetext{b}{In addition to the Absolute RV from \autoref{tab:stellarparam}.}
\tablenotetext{c}{Jitter (per observation) added in quadrature to photometric instrument error.}
\tablenotetext{d}{Where $u_1 + u_2 < 1$, and $u_1 > 0$ according to \cite{kipping_efficient_2013}.}
\tablenotetext{e}{Dilution due to the presence of background stars in \tess{} aperture, not accounted for in the \texttt{eleanor} flux.}
\tablenotetext{f}{We use a Solar flux constant = 1360.8 W/m$^2$, to convert insolation to incident flux.}
\tablenotetext{g}{We assume the planet to be a black body with zero albedo and perfect energy redistribution to estimate the equilibrium temperature. }
\normalsize
\end{deluxetable*}

We perform a joint fit of the photometry (TESS + ground based sources), and the RVs (HPF + NEID) using the \texttt{Python} packge \texttt{exoplanet}, which builds upon \texttt{PyMC3}, the Hamiltonian Monte Carlo (HMC) package \citep{salvatier_probabilistic_2016}. The \texttt{exoplanet} package uses \texttt{starry} \citep{luger_starry_2019, agol_analytic_2020} to model the planetary transits, using the analytical transit models from \cite{mandel_analytic_2002}, and a quadratic limb-darkening law. These limb-darkening priors are implemented in \texttt{exoplanet} using the reparameterization from \cite{kipping_efficient_2013} for uninformative sampling. We fit each phased transit (\autoref{fig:transits}) with separate limb-darkening coefficients. Our likelihood function for the \tess{} photometry includes the GP kernel to model the stellar rotation signal. To account for the long-cadence photometry from \tess{}, \texttt{exoplanet} \citep{foreman-mackey_exoplanet-devexoplanet_2021} oversamples the time series while evaluating the model.

We model the RVs using a standard Keplerian model, where we let the eccentricity float. We include a separate RV offset for each instrument (HPF and NEID), along with a common linear RV trend to account for long term drifts. For the joint modelling of the photometry + RVs, we also include a simple white-noise model in the form of a jitter term that is added in quadrature to the measurement errors from each dataset. 

Using \texttt{scipy.optimize}, we find the initial \textit{maximum a posteriori} (MAP) parameter estimates, which uses the default BFGS algorithm \citep[Broyden–Fletcher–Goldfarb–Shanno algorithm;][]{broyden_convergence_1970, fletcher_new_1970, goldfarb_family_1970, shanno_conditioning_1970}.  These parameter estimates are then used as the initial conditions for parameter estimation using  ``No U-Turn Sampling" \citep[NUTS,][]{hoffman_no-u-turn_2014}, implemented for the HMC sampler \texttt{PyMC3}, where we check for convergence using the Gelman-Rubin statistic \citep[$\hat{\text{R}} \le 1.1$;][]{ford_improving_2006}.


The final derived planet parameters from the joint fit are included in \autoref{tab:planetprop}, with the phased RVs shown in \autoref{fig:RVphase}. We obtain a $\sim 10 \sigma$ mass for TOI-3757 b of \plmass{} M$_\oplus$, and a radius of \plradius{} R$_\oplus$. 

From the joint RV + photometry fit we obtain an eccentricity of 0.14$\pm0.06$, which is consistent with the eccentricity estimate derived from the RBO photometry using the photo-eccentric effect \citep{dawson_photoeccentric_2012}. Here, we compare the stellar density obtained from isochrones and SED fitting in Section \ref{sec:stellar}, to that from the measured transit duration, which is affected by the eccentricity of this orbit. Using this comparison, we measure the eccentricity of the orbit from a joint fit of the TESS and RBO photometry to be 0.12$^{+0.31}_{-0.09}$. This is consistent (to within 1$\sigma$) of the eccentricity derived from the joint RV + photometry fit (\autoref{tab:planetprop}).

\subsection{Upper limit on Helium 10830 \AA~absorption}\label{sec:he10830}
The low bulk density of this planet makes it a promising candidate to detect mass-loss via atmosphere escape. We observed TOI-3757 b during its transit with HPF (described in Section \ref{sec:hpf}), to constrain the absorption in the He 10830 \AA~ triplet due to the planetary exosphere. Not only does HET's fixed-altitude design drastically reduce the number of transits observable with HPF, but also the stellar barycentric velocity shifts the helium triplet feature against the bright hydroxyl sky emission lines. Considering these restrictions, we observed the best available transit of TOI-3757 b with HPF in December 2021.

\begin{figure}[!t] 
\centering
\includegraphics[width=0.48\textwidth]{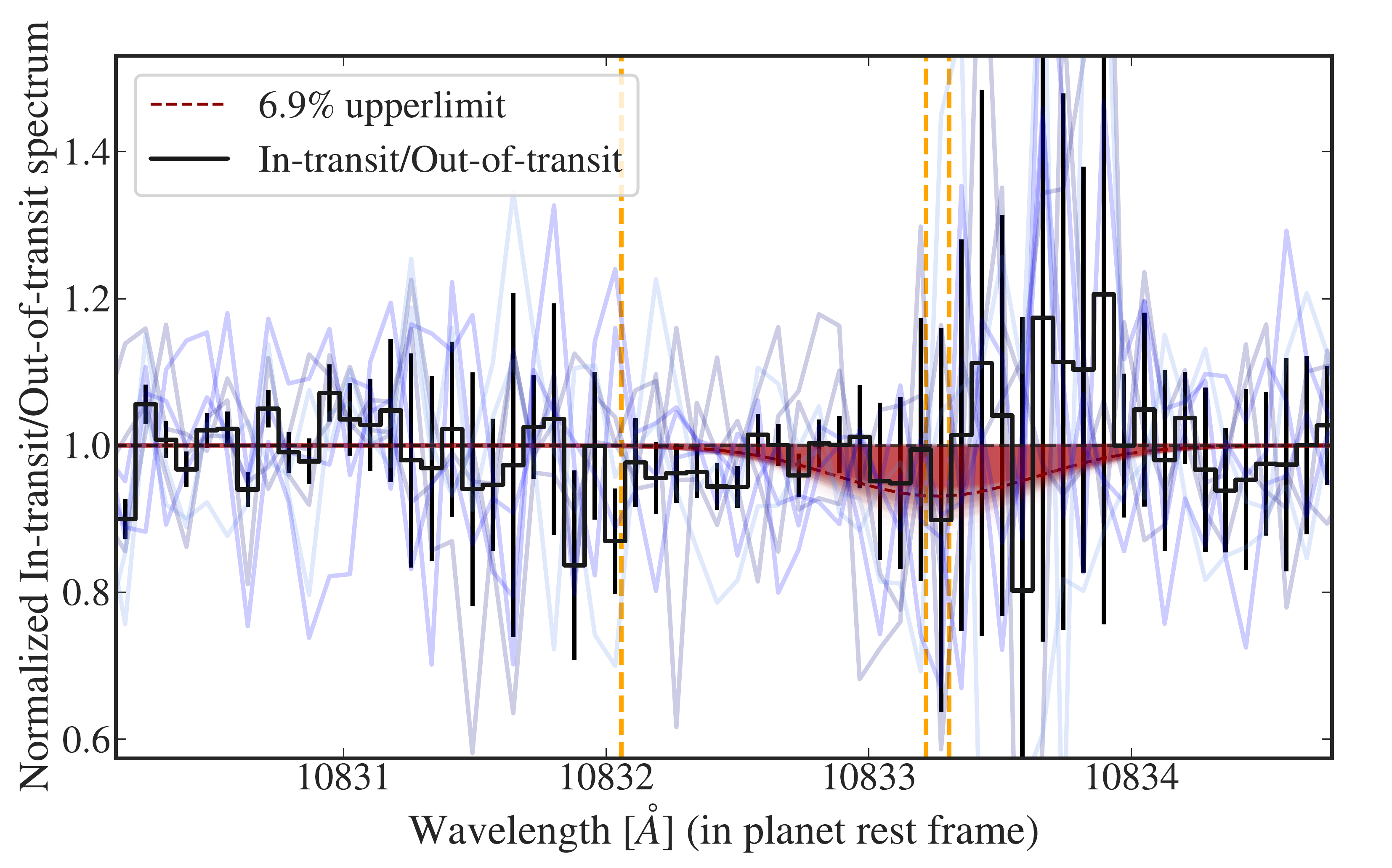} 
\caption{The ratio of the in-transit spectra and out-of-transit TOI-3757 spectra. The blue curves are the three individual ratio spectra from the transit epoch, whereas the black curve is the weighted average of the three. The x-axis shows vacuum wavelength in the planet's rest frame at mid-transit. The rest vacuum wavelengths of the He 10830 \AA, triplet lines in planet’s rest frame are marked by dashed vertical orange lines. We do not detect any significant absorption in the planetary spectra at these wavelengths. The results of our MCMC fit of the strongest doublet lines in the He 10830 \AA~triplet using a Gaussian model of width 0.89 \AA\, are shown by the red curves in the lower panel and the 6.9\% upper-limit is shown by the dashed red curve overlaid on the MCMC results.} \label{fig:he10830}
\end{figure}

The individual spectra were sky subtracted using the simultaneous sky spectra obtained via an adjacent sky fiber \citep{kanodia_overview_2018}. Conservatively, we inflate the errorbars in the pixels that are corrected for sky emission lines to avoid potential systematics. Telluric correction was not performed on this spectra, since the wavelength region of interest for He 10830 \AA~ detection was well outside the telluric absorption lines. We considered all spectra taken 1.9 hours (one full transit duration) away from transit midpoints in the November and December 2021 for the out-of-transit spectra. The individual out-of-transit spectra were then combined by weighted averaging to obtain a high signal-to-noise ratio template spectrum of the star. Individual in-transit spectra were then divided by this template to obtain a set of transmission (or ratio) spectra. These ratio spectra were then aligned to the planet's rest frame (based on the orbit model) and combined by weighted average to obtain the final transmission spectrum shown in Figure \ref{fig:he10830}. No statistically significant He 10830 \AA~ absorption signal was detected. 

For calculating an upper limit, we used a Gaussian absorption line model with an FWHM width of 0.89 \AA. This is the typical width of the reported helium detections in the literature across different planets \footnote{FWHM of He 10830 \AA detections in WASP 69b: 0.72 \AA\, \citep{nortmann_ground-based_2018}, HAT-P-11b:  0.84 \AA\, \citep{allart_spectrally_2018}, HD 189733b : 0.99 \AA\, \citep{salz_detection_2018}, WASP 107b : 0.84 \AA\, \citep{allart_high-resolution_2019, kirk_confirmation_2020}, HD 209458b :  0.44 \AA\, \citep{alonso-floriano_he_2019}, GJ 3470b :  1.2\AA\, \citep{ninan_evidence_2020, palle_he_2020}}. For a circular orbit, we obtain an upper limit on the maximum depth of the line as 6.9\% (with 90\% confidence; see Figure \ref{fig:he10830}). If we consider an eccentricity of 0.14 and argument of periastron to be 130 degrees, the net radial velocity of the planet during the transit is redshifted by 11.2 km/s in comparison to the circular orbit. This places the expected planetary signal on top of a bright hyroxyl sky emission line. Therefore, an eccentric orbit ephemeris precludes us from placing any meaningful upper limit. 
With this caveat, while our upper limit shows the signal is weaker than some of the strongest known signal in exoplanets \citep{allart_high-resolution_2019, kirk_confirmation_2020}, we encourage future spectroscopic and photometric observations to place tighter limits on the presence of He 10830 \AA~ absorption in TOI-3757.

\section{Discussion}\label{sec:discussion}

\subsection{TOI-3757 b: A low density gas giant}

\begin{figure}[!t] 
\centering
\includegraphics[width=0.5\textwidth]{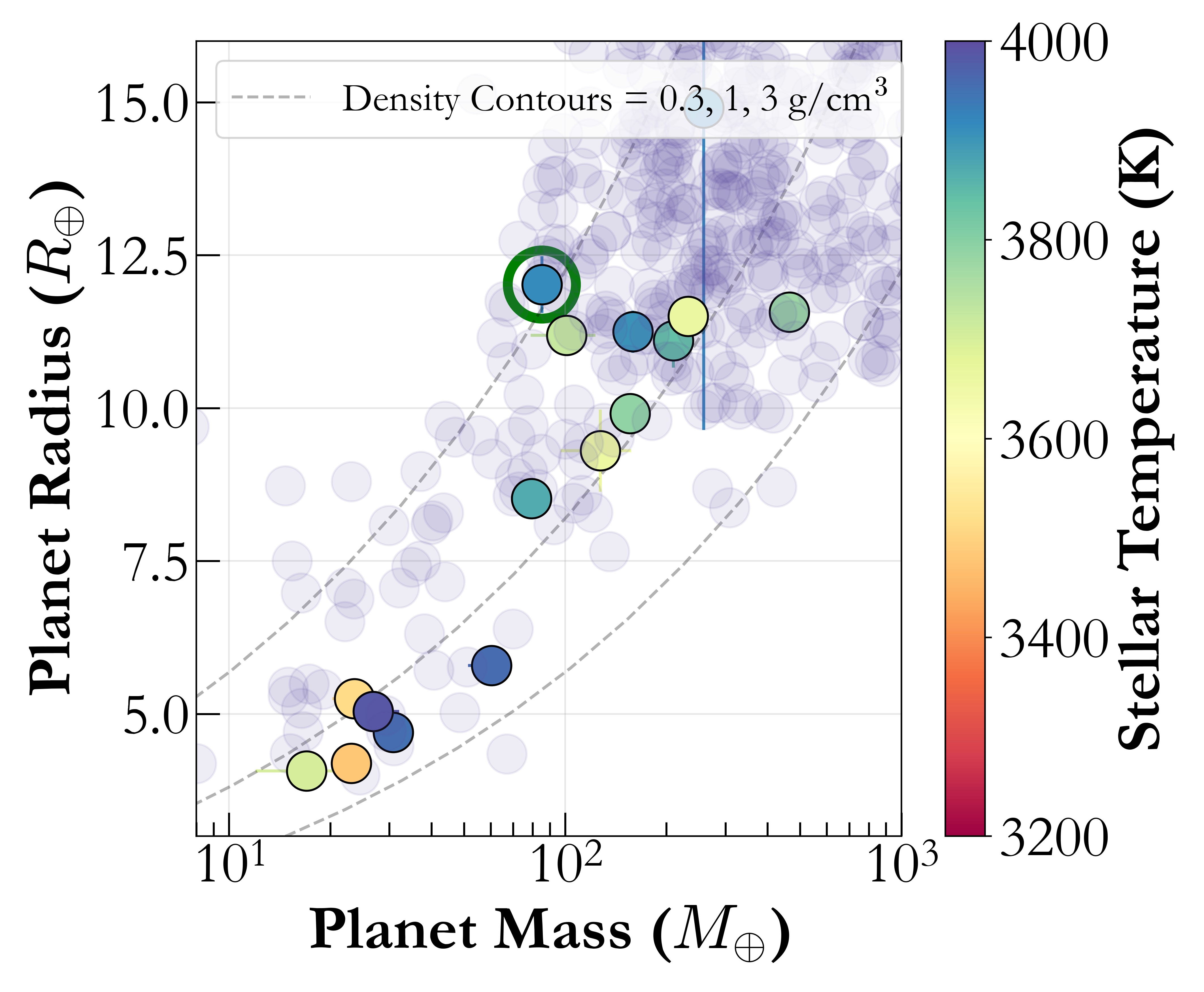}
\caption{We show TOI-3757 b (circled in green) in a mass-radius plane alongside other M dwarf planets (coloured by the \teff{}). We also include planets around FGK stars in the background, along with density contours for 0.3, 1, 3 \gcmcubed{}. TOI-3757 b is the lowest density planet orbiting an M dwarf with precise mass and radius measurements.} \label{fig:MR}
\end{figure}

TOI-3757 b is shown in \autoref{fig:MR} with respect to other M dwarf giant planets ($R_p > 4$ \earthradius{}) with mass measurements at 3$\sigma$ or higher. The data is taken from the NASA Exoplanet Archive \citep{akeson_nasa_2013}, and includes recent M dwarf transiting planets discovered by TESS. While TOI-3757 b is Jovian in size ($\sim 1.05~R_J$), its mass is about one-quarter that of Jupiter ($\sim 0.26~M_J$); due to this TOI-3757 b has the lowest density ($\rho \sim$ 0.27 \gcmcubed{}) hitherto of M dwarf planets with precise mass and radius measurements ($> 3 \sigma$).

We examine different hypotheses to explain the low density (and puffy nature) of TOI-3757 b:

\begin{figure*}[!t]
\fig{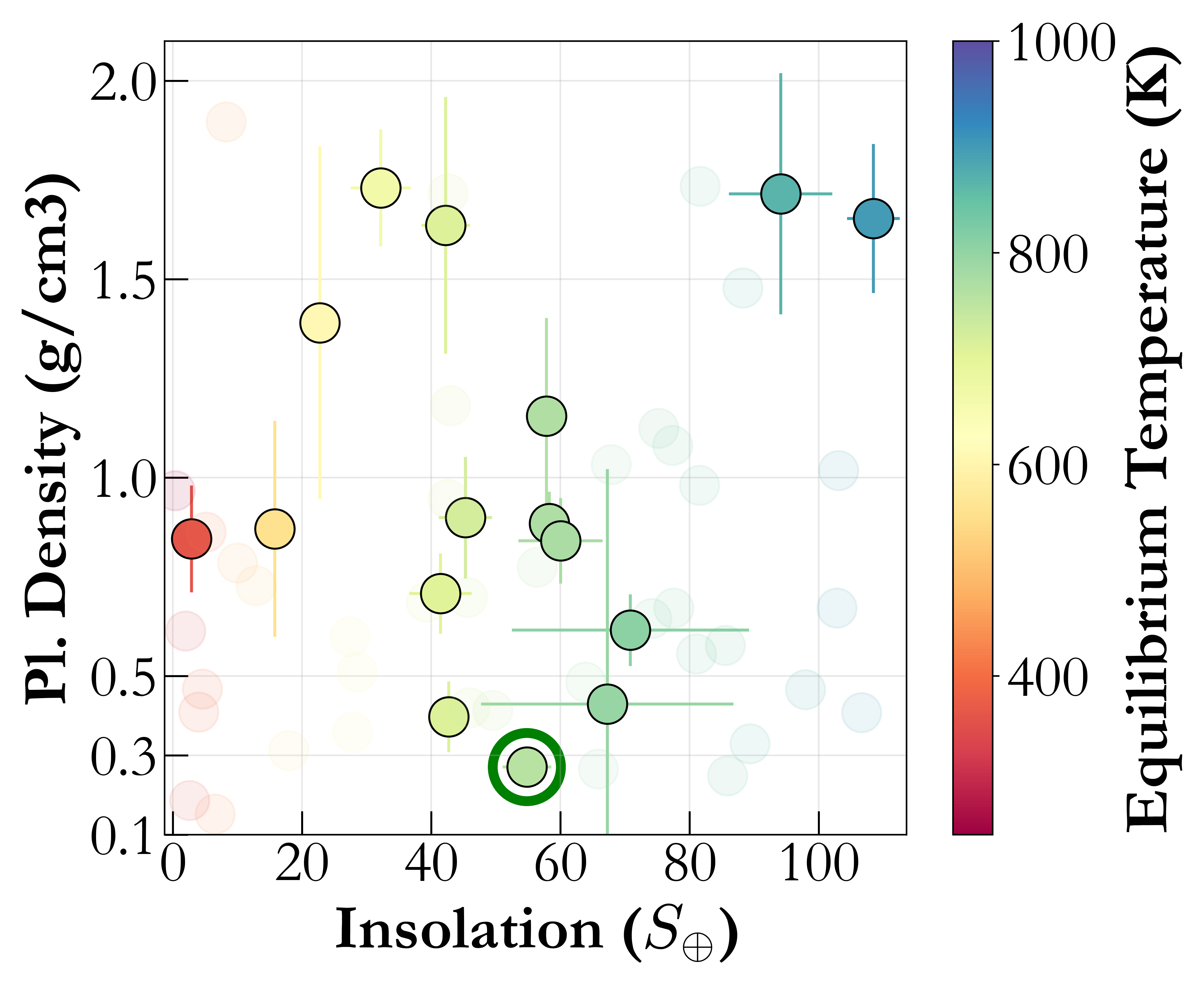}{0.45\textwidth}{\small a) Planet density as a function of insolation flux}    \label{fig:RadiusMass}
\fig{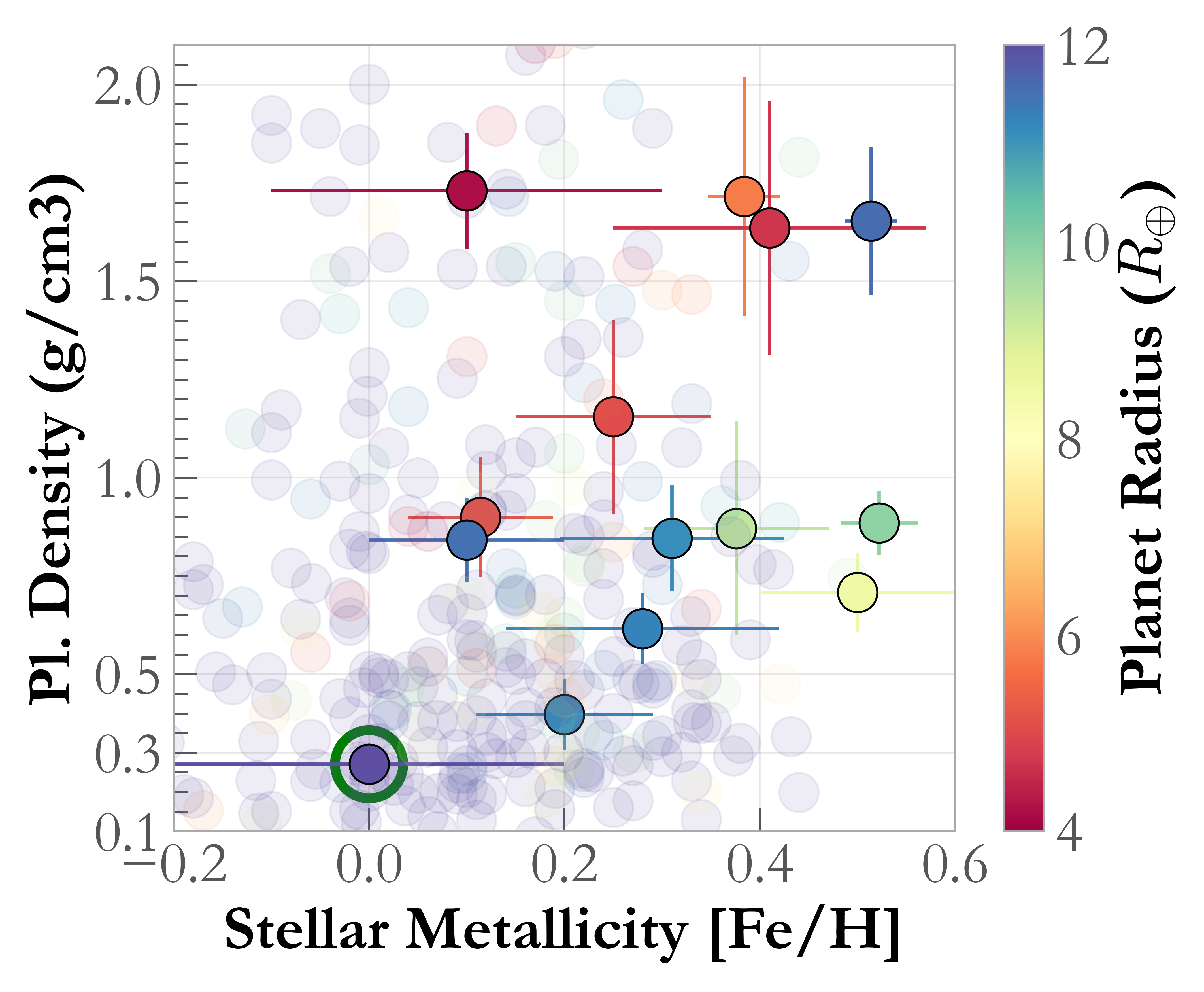}{0.45\textwidth}{ \small b) Planet density as a function of stellar metallicity} \label{fig:RadiusMetallicity}
\caption{\small We show the planetary density of gas giants ($R_p$ $> 4$\earthradius{}) around M dwarfs (solid colours) as a function of stellar insolation and stellar metallicity respectively. In panel \textbf{a)} the markers are colour coded by the equilibrium temperature, whereas panel \textbf{b)} is colour coded by planetary radius. Additionally, TOI-3757 b is highlighted with a green circle. In panel a, we show that TOI-3757 b does not receive an unusually large amount of incident flux from its host star, compared to other similar gas giants. In panel b, we show that the host star has the lowest metallicity of all known gas giants around M dwarfs, which could be a potential explanation of its density. }\label{fig:PlanetDensity}
\end{figure*}

\begin{itemize}
    \item Stellar insolation - The inflated radii of hot-Jupiters\footnote{While formally a hot-Jupiter under the period based range of $1 <$ P $< 10$ days \citep{wright_frequency_2012, wang_occurrence_2015}, TOI-3757 b does not share most of the planetary characteristics of hot-Jupiters, with insolation flux $< 2 \times 10^{5}$ W m$^{-2}$, equilibrium temperature $< 1000$ K \citep{thorngren_mass-metallicity_2016}, and should therefore not be termed such. }, i.e. the so-called radius-anomaly, is primarily explained by the high stellar insolation flux incident on these planets with typical equilibrium temperatures $> 1000$ K \citep[][]{thorngren_bayesian_2018}. Here, a fraction of the absorbed energy from the host star is transported deep into the planetary atmosphere, which then through various mechanisms can cause its inflation\footnote{See \cite{fortney_hot_2021} for a review of inflationary mechanisms of hot-Jupiters.}. However as seen in \autoref{fig:PlanetDensity}a, the insolation flux incident on TOI-3757 b  is not anomalously higher compared to similar M dwarf gas giants. Therefore, we consider it unlikely that the inflated nature of TOI-3757 b is due to stellar insolation.
    
    \item Low metallicity - As indicated in \autoref{fig:PlanetDensity}b, TOI-3757 b has the lowest metallicity of all known gas giants around M dwarfs. It is important to note the caveat here that the M dwarf gas giants form a heterogeneous sample relying on different techniques for metallicity determination\footnote{Refer to \cite{passegger_metallicities_2022} for a detailed discussion on the caveats and challenges associated with estimating M dwarf metallicities.}, and hence this must be interpreted cautiously. While we pursue a detailed discussion on the role of stellar metallicity on the formation of gas giants around M dwarfs in an upcoming manuscript (Kanodia et al. in prep.), we discuss it briefly here. Assuming that the progenitor protoplanetary disk had the same metallicity as the host star, the disk for TOI-3757 is $\sim 0.3$ dex poorer in metallicity compared to the median metallicity of the other transiting gas giants around M dwarfs (\autoref{fig:PlanetDensity}b), albeit with the caveat that our metallicity estimate for the host star TOI-3757 has a 1$\sigma$ error of 0.2 dex. \cite{gan_toi-530b_2021} discuss the metallicity trends seen in the transiting M dwarf Jovian sample, which have a median metallicity of $\sim 0.3$ dex, when the sample is updated to include the Jovian planets that have been confirmed since. This directly influences the formation of the planet in two ways - i) There will be $\sim$ 2x lesser material in the disk than those of comparable gas giants with higher metallicities (10$^{+0.3} \sim 2$) ii)  \cite{yasui_lifetime_2009} suggest that low-metallicity protoplanetary disks have shorter lifetimes and disperse faster (than comparable high-metallicity disks). This would further exacerbate the problem of slow formation timescales for gas-giants around M dwarfs, relative to the disk lifetime \citep{laughlin_core_2004}.
    
    The initiation of runaway gas accretion under the core-accretion theory requires the formation of a rocky core of mass $\sim 10 $\earthmass{} \citep{pollack_formation_1996} and a envelope comparable in mass ($M_{\rm{core}} \sim M_{\rm{env}}$), in a timely manner, before the disk dissipates. Under the hypothesis that the low stellar metallicity is driving the density of TOI-3757 b, due to the two reasons mentioned above, we postulate that for such planets the process of runaway gas accretion did not initiate in a timely enough manner. This would involve a core, which though massive enough for runaway accretion, did not form quickly enough to accrete substantial gas before the disk dissipated. This would explain why TOI-3757 b could not form a planet closer to Jupiter in mass, and is a failed-Jupiter. 
    
    \item Excess internal heat - We explore the possibility that TOI-3757 b's radius is inflated by internal heating (as opposed to external insolation). Multiple studies \citep[e.g.][]{bodenheimer_tidal_2001,burrows.radius.anomalies.2007,lopez.fortney.2014,millholland.2019} show that a planet's radius is influenced by its internal heat - hotter interiors lead to ``puffier'' radii and thus lower densities  compared to cooler interior planets of the same composition. One driver of a hotter interior is simply set by a planet's age; younger planets naturally possess hotter interiors \citep{marley_young_planets}. The lack of a detectable stellar rotation period and SED fit indicate that TOI-3757 is an older star, and therefore we expect its core to have naturally cooled. If TOI-3757 b's inflated radius is due to a hotter interior, it must be heated by some other mechanism. We investigate tidal heating created by orbital eccentricity as a method for inflating the radius of TOI-3757 b. While our eccentric orbit determination for TOI-3757 b is only at the $\sim 2 \sigma$ level with an eccentricity of 0.14$\pm0.06$, we use the framework presented in \citet{leconte.tides} and calculate a tidal-to-irradiation luminosity ratio of $\sim 0.05$ (assuming an eccentricity of 0.14, zero obliquity, and a reduced tidal quality factor --- Q' --- of 10$^{5}$). \cite{millholland_tidal_2020} find that a tidal-to-irradiation luminosity ratio of $>$10$^{-5}$ is required for potential radius inflation due to eccentricity driven tides. With a tidal luminosity ratio of 5\%, TOI-3757 b could be experiencing significant tidal heating from its eccentric orbit. From this power we calculate an interior temperature of 500 K for TOI-3757 b. We also determine that any eccentricity $>$0.001 would lead to some tidal heating in TOI-3757 b and provide interior temperatures $>$200 K. Therefore, it is possible that tides are responsible for at least a part of TOI-3757 b's radius inflation. Quantifying the full impact of tidal heating on the radius inflation however requires a detailed investigation into the interior structure of TOI-3757 b, and a better estimate of eccentricity.
    
    
        
    \item Planetary rings - Recent work by \citet{piro.rings} investigate whether rings could explain the apparent inflated radii of low density super-puffs. For rings to create the appearance of inflated radii due to deeper than expected transits, the rings would need to be at an oblique angle to the planet's orbital plane. If a planet becomes tidally locked to its star however, the ring system must remain in the orbital plane of the planet (no tilt). In this case, we would only observe the edge on portion of the ring during transit and have a negligible impact on the overall transit depth. Using Equation 11 from \citet{piro.rings}, we calculate a synchronous timescale of 5.3 Myrs for TOI-3757 b. If TOI-3757 b did possess rings, they would lie in its orbital/rotation plane and not explain the planet's inflated radii. That said, we are viewing this system from a slightly inclined angle which may tilt the rings into our vantage point. We thus further investigate whether a ring system could remain stable given TOI-3757 b's proximity to its star. \citet{ohta_predicting_2009} note that the Poynting-Robertson drag timescale is extremely short for close-in planets. Using Equation 17 from \citet{ohta_predicting_2009}, we calculate a drag timescale between 0.1 to 5 Myrs for TOI-3757 b dependent on the assumed ring particle densities and sizes. Unless TOI-3757 b also possesses some outside method for stabilizing its ring (such as shepherding moons similar to Saturn), it cannot dynamically maintain a significant ring system.
    

    

\end{itemize}

 Due to the electron degeneracy pressure, Jovian sized objects can span $\sim 100$x in mass, ranging from 0.3 $M_J$ (Saturn mass) to $\sim$ 80 $M_J$ (brown dwarfs and very-low mass stars). Therefore, 2D mass-radius relationships are unable to accurately model Jovian sized gas-giants. The low density for TOI-3757 b also makes it an outlier for mass-radius relationships such as \cite{wolfgang_probabilistic_2016}, \cite{chen_probabilistic_2017}, \cite{ning_predicting_2018} and \cite{kanodia_mass-radius_2019}. Attempts have been made to include orbital period (or insolation) as a third dimension \citep{ma_predicting_2021} in these models to capture the radius anomaly due to host stellar insolation \citep{miller_heavy-element_2011, thorngren_bayesian_2018}. However as shown above, stellar insolation is not the driving force behind the low density for TOI-3757 b, suggesting the need for higher ($> 3$) dimensional analysis including stellar metallicity or eccentricity (among others). We also use the giant planet models from \cite{fortney_planetary_2007} to predict the rocky core mass for TOI-3757 b. While we do not have a precise age estimate, the models assuming a system 1 Gyr old, predict a core mass of $\sim$ 6 \earthmass{}. The models for 10 Gyr \citep[the oldest planetary population modelled by][]{fortney_planetary_2007} do not accommodate objects as low density as TOI-3757 b, however a planet as large as TOI-3757 b would have to have a mass of $\sim 140$ \earthmass{} (which is about 1.6x the mass of TOI-3757 b)  to have a core just $\sim 1$ \earthmass{} in mass. We obtain the same result with the planetary models from \cite{baraffe_structure_2008}, which are also unable to accommodate the low density of TOI-3757 b. 


\subsection{Atmospheric characterization}\label{sec:atmosphere}

\begin{figure}[!t] 
\centering
\includegraphics[width=0.5\textwidth]{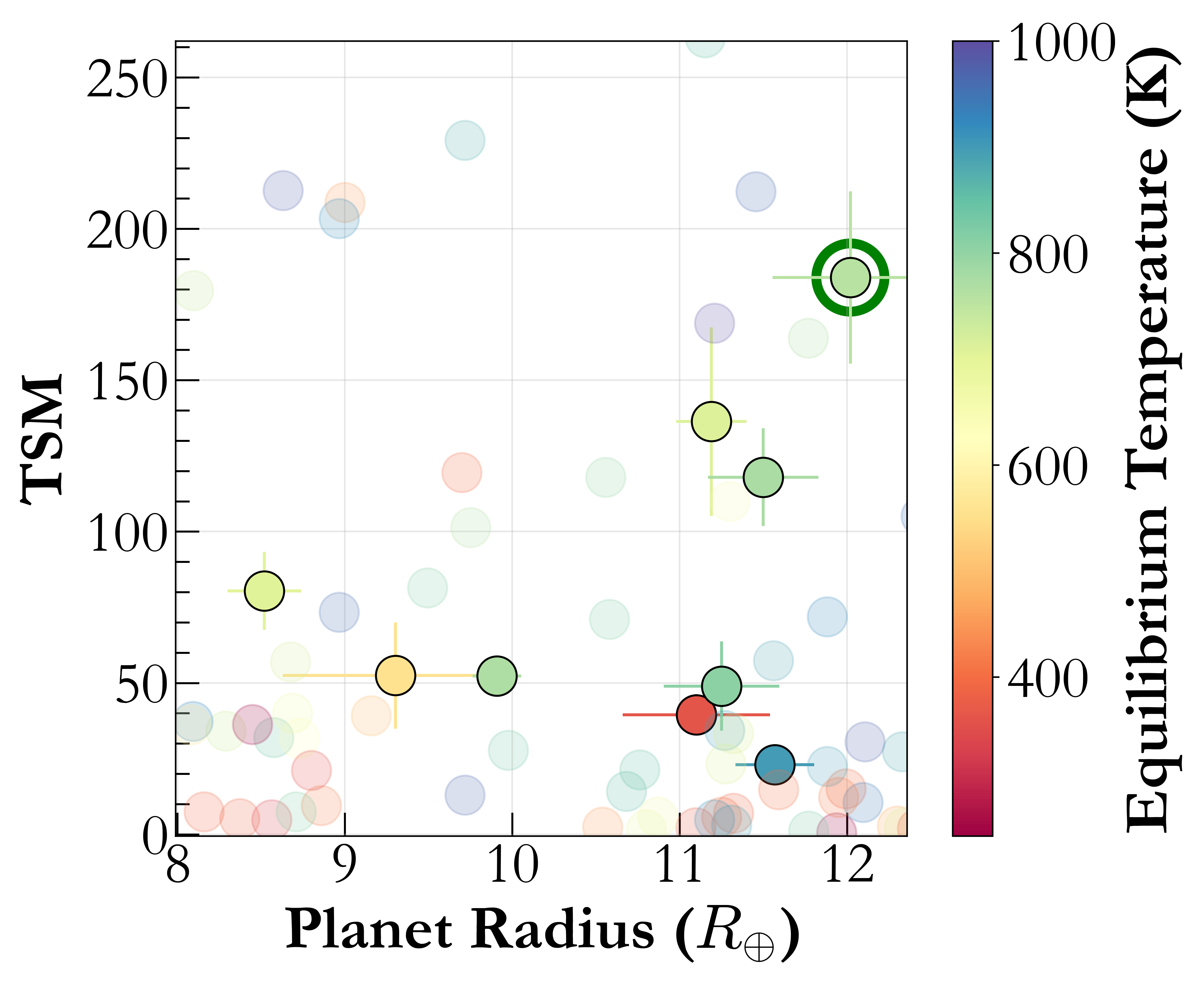}
\caption{We show the TSM for TOI-3757 b (circled in green) as a function of planetary radius, alongside other M dwarf Saturn and Jupiter type planets ($R_p > 6$ \earthradius{}) (coloured by the equilibrium temperature). Similar planets around FGK hosts with equilibrium temperatures $< 1000$ K are shown in the background. TOI-3757 b has the highest TSM among M dwarf gas giants ($R_p > 6$ \earthradius{}) with mass measurements, making it a lucrative target for atmospheric characterization, as discussed in Section \ref{sec:atmosphere}.} \label{fig:TSM}
\end{figure}

TOI-3757 b spans multiple unique regions of parameter space, in terms of its low bulk density, and low host star metallicity (\autoref{tab:planets}). Understanding how TOI-3757 b formed and evolved could provide the context for explaining the low density of this unusual planet. Atmospheric characterization with HST, JWST, and future instruments could provide the necessary information to test our hypotheses regarding TOI-3757 b: is its low density a result of its formational history, or is TOI-3757 b's radius inflated from present day tidal heating? 

Despite the relatively faint host star (J mag $\sim$ 12), the $\sim$ 3\% transit depth and large scale height of $\sim$500 km for TOI-3757 b lend it a large TSM \citep[transmission spectroscopy measurement;][]{kempton_framework_2018} of 190. As shown in \autoref{fig:TSM}, it has one of the highest TSMs of any gas giant ($R_p > 6$ \earthradius{}) with an equilibrium temperature $<$1000 K. Beyond its detectability, TOI-3757's mass (and scale height) is constrained to $>$9$\sigma$ precision allowing us to resolutely analyze its spectrum and removing the planet's mass as a confounding factor during atmospheric retrievals \citep{batalha_precision_2019}.

\begin{figure}[!t] 
\centering
\includegraphics[width=0.5\textwidth]{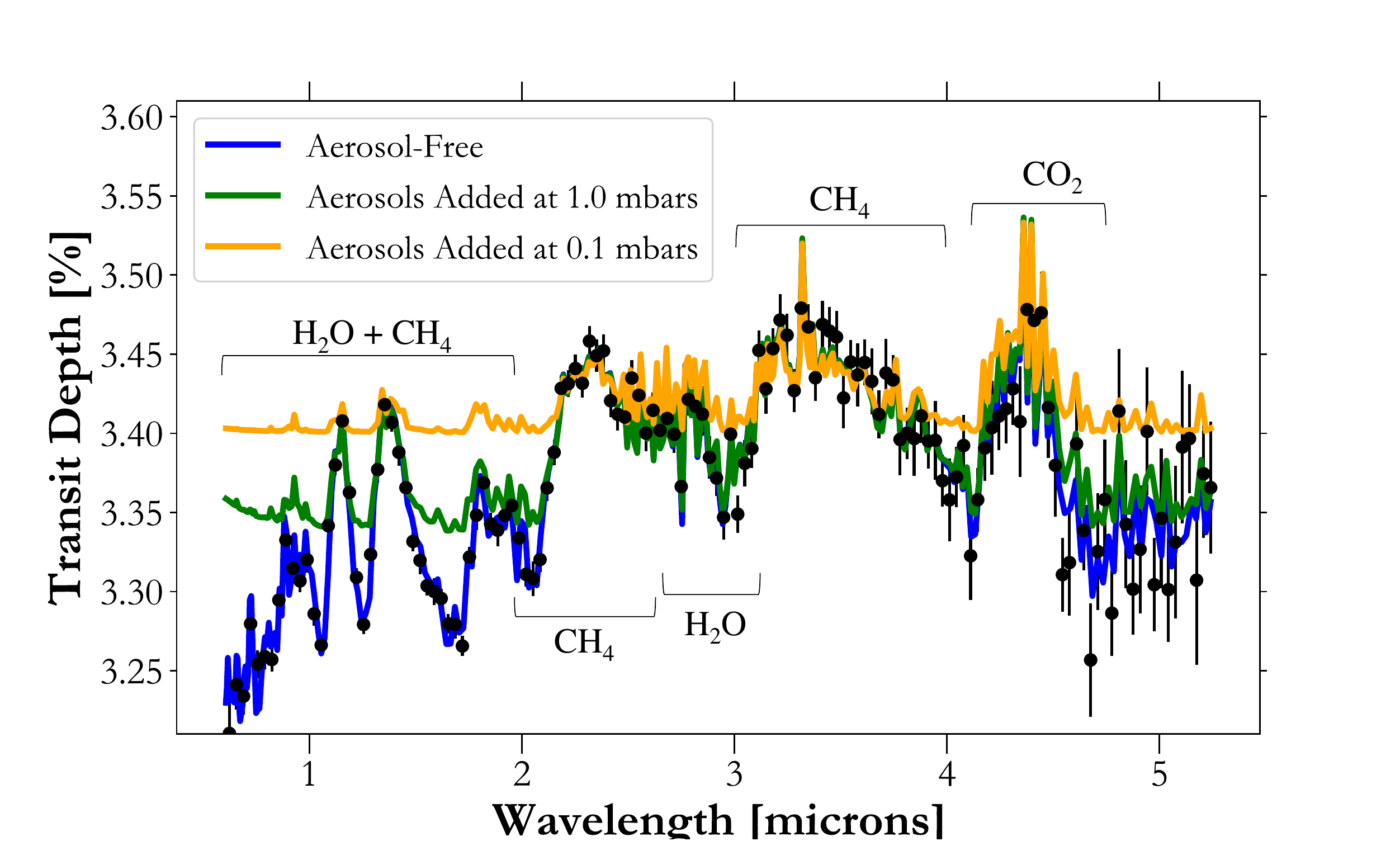}
\caption{We create a simulated JWST/NIRSpec-Prism transmission spectrum of TOI-3757 b (assuming a single transit) using uncertainties calculated with \texttt{PandExo} \citep{batalha_pandexo_2017} assuming a aerosol-free 10$\times$ Solar metallicity atmosphere model \textit{(blue)} generated using \texttt{Exo-Transmit} \citep{kempton_exo-transmit_2017}. We also plot the same underlying composition model plus an opaque aerosol layer at 1 mbars \textit{(green)} and 0.1 mbars \textit{(orange)} for comparison. We assume these aerosols to be grey absorbers (i.e. no wavelength dependence absorption features). Based on \texttt{PandExo} simulations, we find that even with a high-altitude aerosol layer of 0.1 mbars, we should still detect water, methane, and carbon dioxide present in TOI-3757 b's atmosphere with JWST at $\sim 9 \sigma$ compared to a featureless flat spectrum. }
\label{fig:spectra}
\end{figure}

\begin{deluxetable*}{cccccccc}
\tablecaption{List of giant planets with (R$_p > 4$ \earthradius) around M dwarfs, with precise masses ($> 3 \sigma$), and spectroscopic stellar metallicity estimates, as shown in \autoref{fig:PlanetDensity}. Data is taken from the NASA Exoplanet Archive \citep{akeson_nasa_2013}.  \label{tab:planets}}
\centering
\tablehead{\colhead{Host star} & \colhead{Pl. Radius} & \colhead{Pl. Mass} & \colhead{Pl. Density} & \colhead{Orbital Period} & \colhead{\teff{}} & \colhead{Metallicity} & \colhead{Reference} \\
\colhead{} & \colhead{\earthradius{}} & \colhead{\earthmass{}} & \colhead{\gcmcubed{}} & \colhead{days} & \colhead{K} & \colhead{dex} & \colhead{}} 
\startdata
GJ 436   & 4.19$^{+0.11}_{-0.11}$        & 23.1$^{+0.8}_{-0.8}$   & 1.73$^{+0.148}_{-0.148}$     & 2.64388       & 3479$^{+60}_{-60}$   & 0.1$^{+0.2}_{-0.2}$ & \cite{turner_ground-based_2016}   \\
LP 714-47        & 4.7$^{+0.3}_{-0.3}$   & 30.8$^{+1.5}_{-1.5}$   & 1.636$^{+0.323}_{-0.323}$    & 4.05204       & 3950$^{+51}_{-51}$   & 0.41$^{+0.16}_{-0.16}$ & \cite{dreizler_carmenes_2020}       \\
TOI-1728         & 5.04$^{+0.16}_{-0.16}$        & 26.82$^{+5.13}_{-5.44}$        & 1.155$^{+0.247}_{-0.247}$    & 3.492         & 3985$^{+30}_{-30}$   & 0.25$^{+0.1}_{-0.1}$  & \cite{kanodia_toi-1728b_2020} \\
TOI-674          & 5.25$^{+0.17}_{-0.17}$        & 23.6$^{+3.3}_{-3.3}$   & 0.899$^{+0.153}_{-0.153}$    & 1.97714       & 3514$^{+57}_{-57}$   & 0.11$^{+0.07}_{-0.07}$ & \cite{murgas_toi-674b_2021}       \\
TOI-532          & 5.79$^{+0.18}_{-0.19}$        & 60.38$^{+9.1}_{-8.77}$         & 1.715$^{+0.304}_{-0.304}$    & 2.32665       & 3957$^{+69}_{-69}$   & 0.38$^{+0.04}_{-0.04}$  & \cite{kanodia_toi-532b_2021}      \\
TOI-3629         & 8.29$^{+0.22}_{-0.22}$        & 82$^{+6}_{-6}$   & 0.8$^{+0.1}_{-0.1}$    & 3.93655       & 3870$^{+90}_{-90}$   & 0.5$^{+0.1}_{-0.1}$  & \cite{canas_two_2022} \\
HATS-75          & 9.91$^{+0.15}_{-0.15}$        & 156.05$^{+12.4}_{-12.4}$       & 0.884$^{+0.08}_{-0.08}$      & 2.78866       & 3790$^{+5}_{-5}$     & 0.52$^{+0.05}_{-0.03}$    & \cite{jordan_hats-74ab_2021}     \\
Kepler-45        & 10.76$^{+1.23}_{-1.23}$       & 160.49$^{+28.6}_{-28.6}$       & 0.71$^{+0.275}_{-0.275}$     & 2.45524       & 3820$^{+90}_{-90}$   & 0.28$^{+0.14}_{-0.14}$  &  \cite{johnson_characterizing_2012}        \\
HATS-6   & 11.19$^{+0.21}_{-0.21}$       & 101.0$^{+22.0}_{-22.0}$        & 0.397$^{+0.089}_{-0.089}$    & 3.32527       & 3724$^{+18}_{-18}$   & 0.2$^{+0.09}_{-0.09}$ & \cite{hartman_hats-6b_2015} \\
TOI-3714         & 11.32$^{+0.34}_{-0.34}$        & 222.0$^{+10.0}_{-10.0}$        & 0.85$^{+0.08}_{-0.08}$    & 2.15485       & 3660$^{+90}_{-90}$   & 0.1$^{+0.1}_{-0.1}$  & \cite{canas_two_2022}  \\
HATS-74 A        & 11.57$^{+0.24}_{-0.24}$       & 464.03$^{+44.5}_{-44.5}$       & 1.653$^{+0.188}_{-0.188}$    & 1.73186       & 3776$^{+9}_{-9}$     & 0.51$^{+0.03}_{-0.02}$ & \cite{jordan_hats-74ab_2021}        \\
TOI-3757         & 12.02$^{+0.44}_{-0.49}$       & 85.25$^{+8.75}_{-8.67}$        & 0.271$^{+0.041}_{-0.041}$    & 3.43875       & 3913$^{+56}_{-56}$   & 0.0$^{+0.2}_{-0.2}$ & This work  \\
TOI-1899         & 12.89$^{+0.45}_{-0.56}$       & 209.77$^{+22.25}_{-22.25}$     & 0.54$^{+0.08}_{-0.08}$       & 29.02         & 3841$^{+54}_{-45}$   & 0.31$^{+0.11}_{-0.12}$ & \cite{canas_warm_2020}, Lin et al. (in prep.)       \\
\enddata
\tablenotetext{}{We exclude AU Mic b, NGTS-1, and TOI-530 from this table due to lack of spectroscopic metallicity estimates.}
\end{deluxetable*}

With one JWST/NIRSpec-Prism transit of TOI-3757 b, we should easily retrieve an abundance for water, methane, and carbon dioxide in its atmosphere. This combination of molecules provides a constraint on both the overall atmospheric metallicity and the C/O ratio \citep{moses.chemistry,exoplanet.atmosphere.ref}; measurements that inform where in the disk this planet formed \citep{oberg.co.ratio} (\autoref{fig:spectra}). 
We also show in \autoref{fig:spectra} that we should be able to observe molecular absorption features at wavelengths $>$2 microns even with an aerosol layer at 0.1 mbars. While degeneracies between aerosols and atmospheric compositions occur, we demonstrate that TOI-3757 b is a promising target for future JWST observations even if it has high-altitude aerosols present.

Aerosols (condensation clouds or photochemically created hazes) are ubiquitous across all exoplanet atmospheres; flattening or muting features in some while having little apparent effect in the atmosphere of others. By studying Titan in our own solar system, we know UV flux (even a minimal amount) drives the production of photochemical tholin hazes from the photodissociation of methane \citep[e.g.][]{horst.haze.production}. Even though TOI-3757 is an inactive M dwarf, it is likely still producing enough UV flux to create photochemical hazes in TOI-3757 b's atmosphere. We would therefore hypothesize that TOI-3757 b has a hazy atmosphere. However, both \cite{crossfield_trends_2017} and \cite{dymont.trends.2021} found no statistically significant relationship between the UV flux of the star and the water amplitude feature observed by HST/WFC3. Alternatively, \citet{dymont.trends.2021} do detect a tentative correlation between the planet's density and the amplitude indicating that lower density planets may also be more hazy. This trend is greatly influenced by the four super-puff spectra, all of which appear to be featureless \citep{libby-roberts_featureless_2021,kepler79d.featureless,alam.hip41378f}. TOI-3757 b's density is similar to these super-puffs though it is $>$8$\times$ more massive.

TOI-3757 b is instead most similar to the low density sub-Saturns HAT-P-18 b and HAT-P-19 b, both of which orbit early-K dwarf stars \citep{hatp18.hatp19.discovery}. Observed by WFC3, \citet{tsiaras_population_2018} retrieved a clear water abundance for HAT-P-18 b. However, when this spectrum was normalized by the scale height of the atmosphere in \citet{dymont.trends.2021}, the spectrum became near-featureless. It is therefore unclear what we can expect for TOI-3757 b in regards to aerosol formation making it an interesting target for further aerosol studies.

As discussed in \citet{fortney_beyond_2020}, we may also be able to probe the interior temperature of TOI-3757 b by searching for signs of disequilibrium chemistry in the planet's JWST spectrum. The model plotted in \autoref{fig:spectra} assumes chemical equilibrium; however, if TOI-3757 b possesses a hotter interior from tidal heating, methane will appear depleted compared to water which remains unaffected. Moreover, with a strong vertical diffusion coefficient (K$_{zz}$ $>$10$^{4}$) in its atmosphere, ammonia would also be present in TOI-3757 b's JWST spectrum. If neither methane depletion or ammonia is observed, this would also place an upper limit on the interior temperature of TOI-3757, and thus a limit on the tidal heating as a source for radius inflation.

Given TOI-3757 b's low density and extended atmosphere, the planet is potentially experiencing some atmospheric mass loss (even though TOI-3757 is fairly quiescent. Verifying this mass loss is occurring and quantifying the rate would enable us to build a picture of the planet's atmosphere over time. The helium 10830 \AA~ line provides an observable mass loss tracer. As we mentioned in Section~\ref{sec:he10830}, we rule out $>$7\% excess of helium absorption around the planet (assuming a circular orbit); a feature that may be due to a lack of mass loss or the lack of UV stellar flux pumping helium up to its metastable state. We note that helium was recently detected around the low density sub-Saturn HAT-P-18 b with an excess of 0.46\% \citep{helium.hatp18b}. Considering the similarities between the two planets, it is possible that TOI-3757 b maintains an exosphere beneath the precision we achieved with a single transit on HPF. TOI-3757 b would therefore be an interesting target for future helium observations that achieve a higher precision measurement.




\section{Summary}\label{sec:conclusion}
We present the discovery and confirmation of TOI-3757 b, a Jovian sized planet, characterized using a combination of space based photometry from TESS, precise RVs from HPF and NEID, ground based photometric observations from RBO, and speckle imaging from NESSI. With a
planetary radius of \plradius{} \earthradius{}, and mass of \plmass{} \earthmass{}, TOI-3757 b has the lowest bulk density ($\rho =$  0.27$^{+0.05}_{-0.04}$ \gcmcubed{}) of all M dwarf gas giants. Additionally, its host star (TOI-3757) has the lowest stellar metallicity ($\sim$ 0.0 $\pm$ 0.20), of all M dwarfs hosting transiting gas giants. We present different hypotheses to explain the low density of the planet, with the most plausible hypotheses being a formation scenario where its low metallicity is responsible for the delayed on-set of gaseous runaway accretion before the protoplanetary disk dissipated, and an evolution mechanisms where the tidal heating causes the inflation of the planet due to its possibly slightly eccentric orbit (e $\sim$ 0.14 $\pm$ 0.06).

We observed a transit of TOI-3757 b using HPF to place upper limits on helium 10830 \AA~ absorption by the planetary exosphere. Finally, we discuss how the low density (and large scale height) of the planet make it an excellent target for transmission spectroscopy. With just one transit of JWST, we should be able to retrieve the abundance of water and methane in its atmosphere which would enable us to place limits on its C/O ratio. Measuring the levels of methane and ammonia would help constrain the interior temperature of TOI-3757, which can provide insight into potential tidal heating. Additionally, we discuss how TOI-3757 b can also be used to test the correlation between hazy atmospheres and planetary density. 

\section{Acknowledgements}

We thank the anonymous referee for the valuable feedback which has improved the quality of this manuscript. 

The Pennsylvania State University campuses are located on the original homelands of the Erie, Haudenosaunee (Seneca, Cayuga, Onondaga, Oneida, Mohawk, and Tuscarora), Lenape (Delaware Nation, Delaware Tribe, Stockbridge-Munsee), Shawnee (Absentee, Eastern, and Oklahoma), Susquehannock, and Wahzhazhe (Osage) Nations.  As a land grant institution, we acknowledge and honor the traditional caretakers of these lands and strive to understand and model their responsible stewardship. We also acknowledge the longer history of these lands and our place in that history.

These results are based on observations obtained with the Habitable-zone Planet Finder Spectrograph on the HET. We acknowledge support from NSF grants
AST-1006676, AST-1126413, AST-1310885, AST-1310875, AST-1910954, AST-1907622, AST-1909506, ATI 2009889, ATI-2009982, AST-2108512, AST-1907622, and the NASA Astrobiology Institute (NNA09DA76A) in the pursuit of precision radial velocities in the NIR. The HPF team also acknowledges support from the Heising-Simons Foundation via grant 2017-0494.  The Hobby-Eberly Telescope is a joint project of the University of Texas at Austin, the Pennsylvania State University, Ludwig-Maximilians-Universität München, and Georg-August Universität Gottingen. The HET is named in honor of its principal benefactors, William P. Hobby and Robert E. Eberly. The HET collaboration acknowledges the support and resources from the Texas Advanced Computing Center. We thank the Resident astronomers and Telescope Operators at the HET for the skillful execution of our observations with HPF. We would like to acknowledge that the HET is built on Indigenous land. Moreover, we would like to acknowledge and pay our respects to the Carrizo \& Comecrudo, Coahuiltecan, Caddo, Tonkawa, Comanche, Lipan Apache, Alabama-Coushatta, Kickapoo, Tigua Pueblo, and all the American Indian and Indigenous Peoples and communities who have been or have become a part of these lands and territories in Texas, here on Turtle Island.

Data presented herein were obtained at the WIYN Observatory from telescope time allocated to NN-EXPLORE through the scientific partnership of the National Aeronautics and Space Administration, the National Science Foundation, and NOIRLab. This work was supported by a NASA WIYN PI Data Award, administered by the NASA Exoplanet Science Institute. These results are based on observations obtained with NEID on the WIYN 3.5m telescope at KPNO, NSF's NOIRLab under proposals 2021B-0035	(PI: S. Kanodia), 2021B-0435 (PI: S. Kanodia), and 2021B-2015 (PI: S. Mahadevan), managed by the Association of Universities for Research in Astronomy (AURA) under a cooperative agreement with the NSF. This work was performed for the Jet Propulsion Laboratory, California Institute of Technology, sponsored by the United States Government under the Prime Contract 80NM0018D0004 between Caltech and NASA.
WIYN is a joint facility of the University of Wisconsin-Madison, Indiana University, NSF's NOIRLab, the Pennsylvania State University, Purdue University, University of California-Irvine, and the University of Missouri. 
The authors are honored to be permitted to conduct astronomical research on Iolkam Du'ag (Kitt Peak), a mountain with particular significance to the Tohono O'odham. Data presented herein were obtained at the WIYN Observatory from telescope time allocated to NN-EXPLORE through the scientific partnership of NASA, the NSF, and NOIRLab.

Some of the observations in this paper made use of the NN-EXPLORE Exoplanet and Stellar Speckle Imager (NESSI). NESSI was funded by the NASA Exoplanet Exploration Program and the NASA Ames Research Center. NESSI was built at the Ames Research Center by Steve B. Howell, Nic Scott, Elliott P. Horch, and Emmett Quigley.

This work has made use of data from the European Space Agency (ESA) mission Gaia (\url{https://www.cosmos.esa.int/gaia}), processed by the Gaia Data Processing and Analysis Consortium (DPAC, \url{https://www.cosmos.esa.int/web/gaia/dpac/consortium}). Funding for the DPAC has been provided by national institutions, in particular the institutions participating in the Gaia Multilateral Agreement.

Some of the observations in this paper were obtained with the Samuel Oschin Telescope 48-inch and the 60-inch Telescope at the Palomar Observatory as part of the ZTF project. ZTF is supported by the NSF under Grant No. AST-2034437 and a collaboration including Caltech, IPAC, the Weizmann Institute for Science, the Oskar Klein Center at Stockholm University, the University of Maryland, Deutsches Elektronen-Synchrotron and Humboldt University, the TANGO Consortium of Taiwan, the University of Wisconsin at Milwaukee, Trinity College Dublin, Lawrence Livermore National Laboratories, and IN2P3, France. Operations are conducted by COO, IPAC, and UW.

Computations for this research were performed on the Pennsylvania State University’s Institute for Computational and Data Sciences Advanced CyberInfrastructure (ICDS-ACI), including the CyberLAMP cluster supported by NSF grant MRI-1626251.  This content is solely the responsibility of the authors and does not necessarily represent the views of the Institute for Computational and Data Sciences.

The Center for Exoplanets and Habitable Worlds is supported by the Pennsylvania State University, the Eberly College of Science, and the Pennsylvania Space Grant Consortium.

Some of the data presented in this paper were obtained from MAST at STScI. Support for MAST for non-HST data is provided by the NASA Office of Space Science via grant NNX09AF08G and by other grants and contracts.
This work includes data collected by the TESS mission, which are publicly available from MAST. Funding for the TESS mission is provided by the NASA Science Mission directorate. 
This research made use of the (i) NASA Exoplanet Archive, which is operated by Caltech, under contract with NASA under the Exoplanet Exploration Program, (ii) SIMBAD database, operated at CDS, Strasbourg, France, (iii) NASA's Astrophysics Data System Bibliographic Services, and (iv) data from 2MASS, a joint project of the University of Massachusetts and IPAC at Caltech, funded by NASA and the NSF.

This research has made use of the SIMBAD database, operated at CDS, Strasbourg, France, 
and NASA's Astrophysics Data System Bibliographic Services.

This research has made use of the Exoplanet Follow-up Observation Program website, which is operated by the California Institute of Technology, under contract with the National Aeronautics and Space Administration under the Exoplanet Exploration Program

CIC acknowledges support by NASA Headquarters under the NASA Earth and Space Science Fellowship Program through grant 80NSSC18K1114, the Alfred P. Sloan Foundation's Minority Ph.D. Program through grant G-2016-20166039, and the Pennsylvania State University's Bunton-Waller program.

SK would like to acknowledge E.H. Mason for help with this project.

This research made use of \textsf{exoplanet} \citep{foreman-mackey_exoplanet-devexoplanet_2021, foreman-mackey_exoplanet_2021} and its
dependencies \citep{agol_analytic_2020, foreman-mackey_fast_2017, foreman-mackey_scalable_2018, kumar_arviz_2019, robitaille_astropy_2013, astropy_collaboration_astropy_2018, kipping_efficient_2013, luger_starry_2019, the_theano_development_team_theano_2016, salvatier_probabilistic_2016}

\facilities{\gaia{}, HET (HPF), WIYN(NEID), \tess{}, RBO, Exoplanet Archive}

\software{
\texttt{ArviZ} \citep{kumar_arviz_2019}, 
AstroImageJ \citep{collins_astroimagej_2017}, 
\texttt{astroquery} \citep{ginsburg_astroquery_2019}, 
\texttt{astropy} \citep{robitaille_astropy_2013, astropy_collaboration_astropy_2018},
\texttt{barycorrpy} \citep{kanodia_python_2018}, 
\texttt{celerite2} \citep{foreman-mackey_fast_2017, foreman-mackey_scalable_2018}
\texttt{exoplanet} \citep{foreman-mackey_exoplanet-devexoplanet_2021, foreman-mackey_exoplanet_2021},
\texttt{HxRGproc} \citep{ninan_habitable-zone_2018},
\texttt{ipython} \citep{perez_ipython_2007},
\texttt{lightkurve} \citep{lightkurve_collaboration_lightkurve_2018},
\texttt{matplotlib} \citep{hunter_matplotlib_2007},
\texttt{MRExo} \citep{kanodia_mass-radius_2019},
\texttt{numpy} \citep{oliphant_numpy_2006},
\texttt{pandas} \citep{mckinney_data_2010},
\texttt{PyMC3} \citep{salvatier_probabilistic_2016},
\texttt{scipy} \citep{oliphant_python_2007, virtanen_scipy_2020},
\texttt{SERVAL} \citep{zechmeister_spectrum_2018},
\texttt{starry} \citep{luger_starry_2019, agol_analytic_2020},
\texttt{Theano} \citep{the_theano_development_team_theano_2016}.
}
\bibliography{references, jlr_references}

\end{document}